\documentclass[traditabstract]{aa}
\usepackage{graphicx}
\usepackage{txfonts}
\usepackage{longtable}
\usepackage[authoryear]{natbib}
\bibpunct{(}{)}{;}{a}{}{,}

\def\ltsima{$\; \buildrel < \over \sim \;$}
\def\simlt{\lower.5ex\hbox{\ltsima}}
\def\gtsima{$\; \buildrel > \over \sim \;$}
\def\simgt{\lower.5ex\hbox{\gtsima}}

\begin{document}
   \title{Chemical abundances in the nucleus of the Sagittarius dwarf spheroidal galaxy}

   \author{A. Mucciarelli\inst{1,2}, M. Bellazzini\inst{2}, R. Ibata\inst{3}, D. Romano\inst{2}, S.C. Chapman, \inst{4}, L. Monaco\inst{5}}
         
      \offprints{A. Mucciarelli}

   \institute{Dipartimento di Fisica e Astronomia, Universit\`a degli Studi di Bologna, 
               via Gobetti 93/2, I-40129 Bologna, Italy;
             \email{alessio.mucciarelli2@unibo.it} 
             \and
  			INAF - Osservatorio Astronomico di Bologna,
              Via Gobetti 93/3, I-40129, Bologna, Italy
              \and
             Observatoire astronomique de Strasbourg, Universit\'e de Strasbourg, CNRS, UMR 7550, 11 rue 
             de l'Universit\'e, F-67000 Strasbourg, France 
             \and
             Department of Physics and Atmospheric Science, Dalhousie University, Halifax, NS B3H 4R2, Canada 
             \and
             Departamento de Ciencias Fisicas, Universidad Andres Bello, Fernandez Concha 700, Las Condes, Santiago, Chile}

     \authorrunning{A. Mucciarelli et al.}
   \titlerunning{Chemical abundances in the nucleus of the Sagittarius dwarf spheroidal galaxy}

   \date{Submitted to A\&A }

\abstract{We present Iron, Magnesium, Calcium, and Titanium abundances for 235 stars in the central region 
of the Sagittarius dwarf spheroidal galaxy (within $9.0\arcmin\simeq 70$~pc from the center) from  
medium-resolution Keck/DEIMOS spectra. All the considered stars belong to the massive globular cluster M~54 
or to the central nucleus of the galaxy (Sgr,N). In particular we provide abundances for 109 stars with [Fe/H]$\ge -1.0$, more 
than doubling the available sample of spectroscopic metallicity and $\alpha$-elements abundance estimates for Sgr~dSph 
stars in this metallicity regime. 
We find for the first time a metallicity gradient in the Sgr,N population, whose peak iron abundance goes from [Fe/H]=$-0.38$ 
for $R\le 2.5\arcmin$ to  [Fe/H]=$-0.57$ for $5.0<R\le 9.0\arcmin$. On the other hand the trends of [Mg/Fe], [Ca/Fe], and 
[Ti/Fe] with [Fe/H] are the same over the entire region explored by our study. We reproduce the observed chemical patterns 
of the Sagittarius dwarf spheroidal as a whole with a chemical evolution model implying a high mass progenitor ($M_{DM}=6\times 10^{10}~M_{\sun}$) and a significant event 
of mass-stripping occurred a few Gyr ago, presumably starting at the first peri-Galactic passage after infall.}


\maketitle
%

\section{Introduction}
\label{intro}

The Sagittarius dwarf spheroidal galaxy \citep[Sgr~dSph,][]{iba04} is the most emblematic case of a Galactic satellite that is 
being disrupted by the tidal field of the Milky Way. The main body of the system is a large 
\citep[half-light radius $r_h=2.6$~kpc,][]{m03} low-surface brightness gas-less elongated spheroid. It is located behind the Galactic bulge, 
as seen from the Sun, and $\sim 6.5$~kpc below the Galactic plane, lying at a distance of $D=26.3\pm 1.8$~kpc from us \citep{l_tip}. On the other hand, the two arms of 
its tidal stream are traced all over the sky, out to $D\simeq 100$~kpc \citep[see][and references therein]{m03,bel14}. 
The process of disruption of Sgr~dSph is contributing to the build-up of the Galactic halo in terms of dark matter, stars, and 
globular clusters \citep[see, e.g.,][]{m03,b03,l_hb,lm10,gibbons,deboer}. 

According to the compilation by \citet{mc12}, the present-day stellar mass and dynamical mass (within $r_h$) of the main body are 
$M_{\star}\sim 2\times 10^7~M_{\sun}$ \citep{iba94} and $M_{dyn}\sim 2\times 10^8~M_{\sun}$ \citep{grc09}, respectively, but 
there is a consensus that the mass of the progenitor must have been 
a factor of ten larger \citep{nied12,deboer,gibbons}. 
The stellar population of the main body is dominated by an 
old/intermediate-age (mean age$\ga 5$~Gyr) relatively metal-rich ($\langle {\rm [Fe/H]}\rangle \simeq -0.5$) population 
\citep[][and references therein]{b06b,siegel}, with an old and metal-poor component contributing $\sim 10$ per cent 
to the stellar budget \citep{l_hb,hama}.

In spite of being extended over several kpc, Sgr~dSph hosts an over-density of stars on the $\la 100$~pc scale lying straight 
at its center (see Fig.~\ref{mappat}), with all the characteristics of the stellar nuclei very frequently found in dwarf ellipticals 
as well as in galaxies of other morphological types \citep[see][B08 hereafter, references and discussion therein]{l_nuc,b08}. 
The massive and metal-poor ([Fe/H]=$-1.56$, \citealt{c10b}) globular cluster M~54 (NGC~6715) coincides with this nucleus 
(Sgr,N hereafter) in position and radial velocity. B08  showed that the two systems have different surface density profiles 
and, above all, very different velocity dispersion profiles, supporting the idea that M~54 formed independently from the 
metal-rich nucleus and was brought to its current position by dynamical friction \citep[see also][]{iba09}. The viability of the scenario was demonstrated 
both analytically \citep{l_nuc} and with N-body simulations (B08), and it found independent support from chemical tagging arguments \citep{c10a}.

The chemical composition of stars in Sgr~dSph is of great interest and indeed several spectroscopic studies addressed 
this problem, e.g., \citet[][SM02 hereafter]{sm02}, \citet{bo00,bo04}, \citet[][M05 hereafter]{mo05}, \citet[][S07]{sbo07}, 
that re-analysed also Bonifacio et al.'s samples, \citet{chou}, \citet[][C10a and C10b, respectively]{c10a,c10b}, \citet[][MWM13]{mcw13}. 
In particular, MWM13 present a comprehensive view of all the iron and $\alpha$-elements abundances available in the literature 
for the main body of the galaxy. Surprisingly enough they amount to less than 70 stars, the large majority having [Fe/H]$\ge -1.0$. 
Hence we are far from having a satisfactory sampling of the metallicity distribution and abundance patterns in this crucial system. 

To make a significant step ahead in this direction we present here the elemental abundances of Iron, Magnesium, Calcium, and Titanium
for a large and homogeneous subsample of the M~54 and Sgr,N stars studied in B08 using Keck/DEIMOS spectra. 
These authors estimated the metallicity of their stars using the Calcium triplet technique, with the main purpose of classifying 
stars for the analysis of the kinematics of the system, that was the final goal of their study. Conversely, here we analyse the chemical 
abundances of stars derived via synthetic spectroscopy on individual atomic transition lines. Since the main abundance patterns and trends in M~54 have been already studied and discussed in detail by C10b we focus our analysis on the Sgr population.

The plan of the paper is the following. In Section~\ref{data} we describe the sample and the data analysis and in Section~\ref{compa}
we compare our results with those available in the literature. Section~\ref{chem} is devoted to the analysis of 
the observed abundance distributions and chemical patterns, also by means of chemical evolution models. The main results 
of the analysis are summarised and discussed in Section~\ref{conc}.

\section{Data and abundance measures}
\label{data}

\subsection{Observational dataset}
\label{obs}

\begin{figure}
   \centering
   \includegraphics[width=\columnwidth]{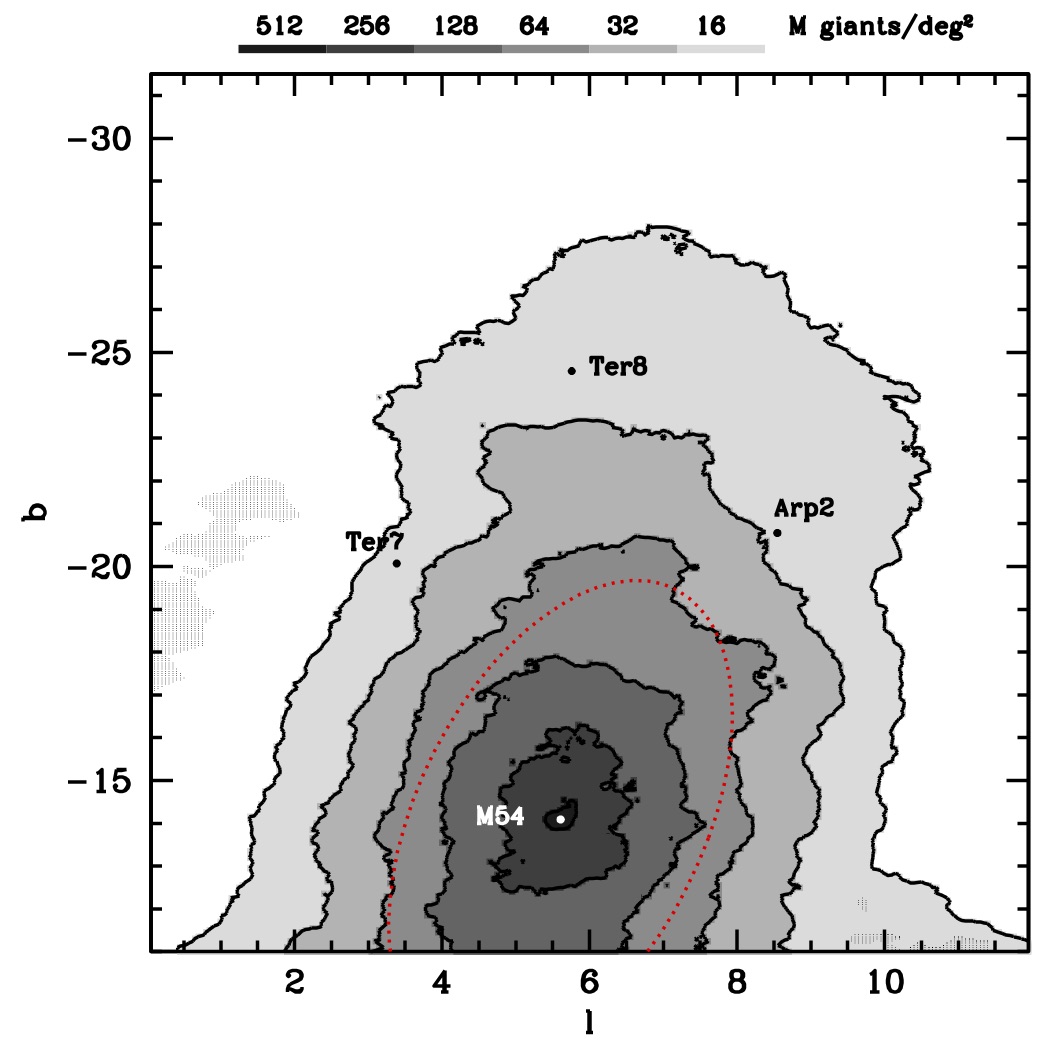}
     \caption{Surface density map of the southern half of the main body of Sgr~dSph as traced by M giants from 2MASS \citep{2mass}, selected according to \citet{m03}. The light-shaded areas have E(B-V)$>0.3$ according to the extinction maps by \citet{sfd98}. 
     The globular clusters associated to Sgr~dSph are marked and labelled. The dashed ellipse is the half-light contour 
     of the best-fit \citet{k62} model by \citet{m03}; the half-light radius along the major axis is taken from \citet{mc12}. For $b\ge -13.0\degr$ the contamination by foreground Galactic M-giants and the high extinction prevent a reliable tracing of the Sgr density, affecting the shape of the outer iso-density contours.
     At the distance of Sgr~dSph \citep[D=26.3~kpc][]{l_tip} one  degree corresponds to 459~pc.}
        \label{mappat}
    \end{figure}

\begin{figure}
   \centering
   \includegraphics[width=\columnwidth]{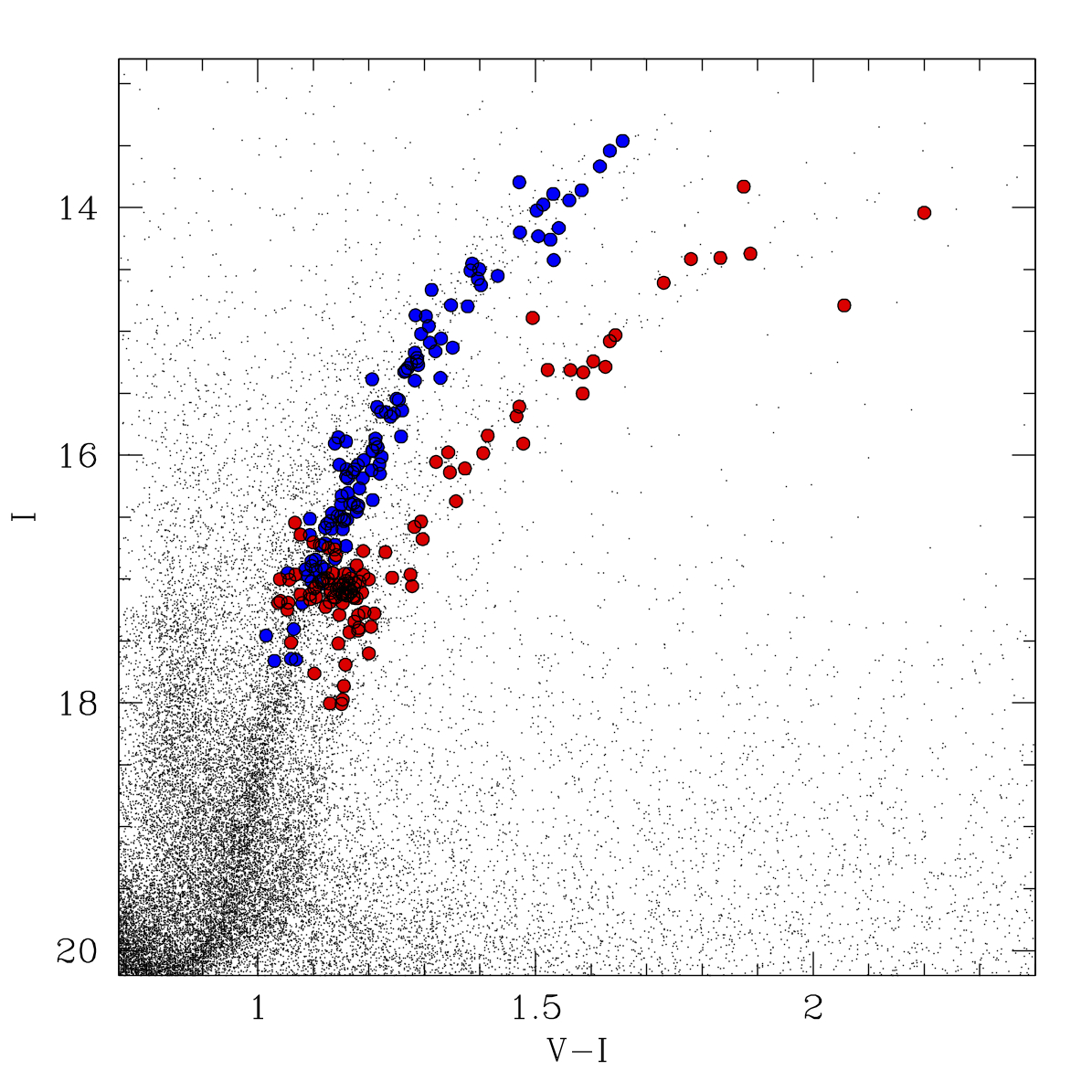}
     \caption{Color-magnitude diagram of the inner 9.0' of M54 \citep{monaco02} with superimposed the position 
     of the Keck/DEIMOS spectroscopic targets: red circles are the stars with [Fe/H]$>$-1.0 dex, 
     blue circles those with [Fe/H]$<$--1.0 dex.}
        \label{cmd}
    \end{figure}

The spectroscopic data analysed in this work are part of the dataset already discussed in B08. 
The bulk of the B08 sample was constituted by Keck/DEIMOS spectra at ${\tt R}=\frac{\lambda}{\Delta \lambda}\simeq 6600$ 
fully analogous to those used by \citet{Mu12} to derive abundances of a few chemical elements in the globular cluster NGC~2419 
with spectral synthesis of selected atomic lines. 
The original sample of B08 used both {\sl slits} and {\sl holes} DEIMOS spectra. 
In the present work, we limit the analysis to the DEIMOS {\sl slits} spectra,
because the DEIMOS {\sl holes} spectra cover a smaller spectral range 
that prevents to measure the Mg abundances and leads to a significant decrease of the number 
of available Fe lines (down to only 5 instead of $\sim$15).
We perform the same analysis as in \citet{Mu12} on 235 stars observed with Keck/DEIMOS whose membership to M54 or Sgr,N has been established 
by B08 and selected to have signal-to-noise ratio (SNR) larger than $\sim$20-30 (up to $\sim$120-130) in the spectral region $\sim8400-8700$ \AA\ . 
This sample is confined to the nuclear region of Sgr. All the stars are within $9.0\arcmin \simeq 70$~pc from 
the center of the system, they are all enclosed within the innermost density contour of Fig.~\ref{mappat}. 
The position on the color-magnitude diagram (CMD) of the spectroscopic targets analysed in this paper is shown in Fig.~\ref{cmd}.
We limited the analysis to this subsample of the B08 data to keep the observational material and the analysis as homogeneous as possible, 
to have the most self-consistent scale of relative abundances.

The original B08 sample was selected to minimise the contamination by Galactic foreground stars by choosing 
targets lying close to the main branches of M~54 and Sgr,N on the CMD. This implies that 
the resulting metallicity distribution for Sgr,N is biased against metal-poor stars not only because in this central 
region it is overwhelmed by M~54 stars but also because stars with intermediate colours between the blue Red Giant Branch (RGB) 
of M~54 and the main red RGB of Sgr were not included in the 
sample\footnote{We note that any attempt to obtain an unbiased metallicity distribution in Sgr would require to sample a 
region far from M~54 or other Sgr globulars and would imply to take the spectra of a large number of Galactic foreground 
stars having color and magnitude in the range spanned by Sgr red giants.}.  
On the other hand our sample is fully adequate to study the peak of the metal-rich population and to trace the trends of 
abundance ratios with [Fe/H].

\subsection{Atmospheric parameters}

Effective temperatures ($T_{eff}$) and surface gravities (log~g) have been 
derived from the photometric catalog by \citet{monaco02}. 
In particular, $T_{eff}$ have been obtained from the $(V-I)_{0}-T_{eff}$ 
calibration provided by \citet{alonso99} and adopting the color excess 
E(B-V)=~0.14$\pm$0.03 mag \citep{ls00}. Because the \citet{alonso99} calibration 
used the Johnson I-filter, the Cousin I-band magnitudes by \citet{l_tip} 
have been transformed into the Johnson photometric system adopting the transformation provided by 
\citet{bessell79}. Internal uncertainties in $T_{eff}$ are estimated 
including errors in photometric data and reddening and they are of the order 
of $\sim$60-100 K.

Surface gravities have been calculated through the Stefan-Boltzmann 
relation and adopting the photometric $T_{eff}$ described above, the bolometric 
corrections by \citet{alonso99}, a stellar mass of 0.8 $M_{\odot}$ and 
a true distance modulus $(m-M)_0$=~17.10$\pm$0.15 mag \citep{l_tip}. 
Propagating the uncertainties in $T_{eff}$, E(B-V), mass and distance modulus 
we obtain uncertainties in log~g of $\pm$0.1.

Microturbulent velocities ($v_t$) cannot be derived directly from the spectra, 
as usually done with high-resolution spectra, because the spectral resolution 
of the Keck/DEIMOS spectra prevents the measurement of weak Fe~I lines. 
For this work, the log~g--$v_t$ calibration provided by \citet{kirby09} 
has been adopted. Note that the formal errors in log~g lead to uncertainties 
in $v_t$ of only $\pm$0.02-0.03 km/s. Because the used lines are strong enough 
to be sensitive to the velocity fields, we assumed a conservative uncertainty 
of 0.2 km/s.

\subsection{Abundance analysis}

Chemical abundances have been derived with our own code {\tt SALVADOR} through a $\chi^2$-minimization 
between the observed line and a grid of synthetic spectra calculated with 
the appropriate atmospheric parameters and varying only the abundance of the 
species under scrutiny. The synthetic spectra have been calculated 
using the code SYNTHE\footnote{http://wwwuser.oats.inaf.it/castelli/sources/synthe.html} 
and including all the transitions included in the Kurucz/Castelli linelists.
Model atmospheres have been calculated with the last version of the ATLAS9
code\footnote{http://wwwuser.oats.inaf.it/castelli/sources/atlas9codes.html}.
The transitions to be used for the chemical analysis have been selected 
according to suitable synthetic spectra calculated over the entire 
wavelength range and convolved with a Gaussian profile to reproduce 
the spectral resolution of the Keck/DEIMOS spectra. 
We picked only unblended transitions using the atomic data listed 
in the Kurucz/Castelli linelists\footnote{http://wwwuser.oats.inaf.it/castelli/linelists.html}.
We included in our linelist 15 Fe~I lines, 4 Ti~I lines, the 
Mg~I line at 8806.7 \AA\ , the two strongest CaT lines at 8552 and 8662 \AA\ . 
For each transition (except the Ca~II lines) the $\chi^2$-minimization 
has been performed within a fitting-window of $\sim$4-5 \AA\ . 
Only for the Ca~II lines we adopted a different strategy: 
following \citet{Mu12}, the Ca abundance has been derived 
by fitting only the wings of these lines and disregarding the core. 
In fact, the core of the Ca~II lines, forms in the outermost photospheric layers and it 
is sensitive to the  inadequacies of the 1-dimensional 
model atmospheres and it suffers
from severe departures from local thermodynamical equilibrium.

All the observed spectra have been previously normalized by fitting the entire spectrum 
with a cubic spline function. The continuum location has been refined by using 
pre-selected spectral windows representative of the continuum (i.e. without spectral features): 
a cubic spline has been calculated considering the median values of the continuum windows 
and then used to normalize, in a self-consistent way, 
both synthetic and observed spectra.

Because of the large range of metallicity covered by the entire sample 
(including both metal-poor M54 stars and metal-rich Sgr,N stars), a unique list of 
continuum windows cannot be suitable for all the stars. The entire analysis procedure 
has been iteratively performed: a first analysis has been done adopting a 
preliminary set of windows and then refined with an appropriate list of 
continuum windows selected according to the stellar metallicity.

The procedure of normalization can be critical at the DEIMOS spectral resolution, 
especially for the most metal-rich stars, where the blanketing conditions 
can make difficult an appropriate location of the continuum level. 
In order to check the stability of our abundances against the method to normalize 
the spectra, we repeat the analysis for a subsample of spectra adopting the 
approach originally described by \citet{shetrone09} and already implemented in 
{\tt SALVADOR} \citep[see][]{Mu12}. In this approach, the observed spectrum is divided by the spline 
obtained fitting the residuals between the observed spectrum and the best-fit 
synthetic spectrum obtained in a first iteration. The procedure is repeated 
if the derived abundance changes significantly.
We analysed some tens of stars picked in order to sample different metallicity and 
SNR. For bright, metal-poor stars the absolute differences in the derived abundances 
between the two methods are smaller than 0.02 dex, while 
for the bright, metal-rich stars do not exceed 0.04 dex.
For faint stars, with low SNR, the differences between the two approaches remain 
smaller than 0.1 dex. However, these abundance differences are, on average, 
compatible with zero, pointing out that the two approaches are substantially equivalent.
The use of the method of \citet{shetrone09} does not change significantly the derived 
metallicity distribution of Sgr,N and our conclusions.

Uncertainties in the chemical abundances 
have been estimated by taking into account two main sources of error: 
\begin{enumerate}
\item ~Uncertainties arising from the fitting procedure that have been 
estimated by using Montecarlo simulations. For each spectral line, 300 simulated 
spectra have been created by re-sampling the best-fit synthetic spectrum at the 
pixel-scale of the observed spectrum and then 
by adding Poissonian noise in order to mimic the measured SNR per pixel. 
This set of simulated noisy spectra 
has been re-analysed with the same procedure used for real spectral features 
and the standard deviation of the derived abundance distribution is taken as 
the uncertainty. For the lines observed in the brightest stars 
(I$\simeq$14.0 and SNR$\sim$130), the fitting uncertainty is about $\pm$0.03-0.04 dex, 
increasing up to $\sim$0.2 dex for the faintest stars (I$\simeq$18.0 and SNR$\sim$20-30). 
Only for the Ca lines, the error associated with the photon noise is smaller 
(reaching $\pm$0.05 dex for the faintest stars) because of the large number 
of pixels used in the fitting procedure;
\item ~Uncertainties arising from the atmospheric parameters that 
have been estimated by repeating the analysis varying each time 
one atmospheric parameter by the corresponding uncertainty. 
The used lines are mainly sensitive to $T_{eff}$ and $v_t$, while 
marginally sensitive to log~g.
A variation in log~g of $\pm$0.1 leads to a variation in [Fe/H] of $\pm$0.02-0.03 dex 
and virtually a null variation in the other abundance ratios. 
On the other hand, a variation in $T_{eff}$ of $\pm$100 K leads to 
typical changes of $\pm$0.05-0.07 dex in [Fe/H], $\pm$0.10 dex in [Mg/Fe], 
$\mp$0.05 dex in [Ca/Fe] and up to $\pm$0.15-0.20 dex in [Ti/Fe] (as 
the Ti~I lines used in this work are very sensitive to changes in $T_{eff}$). 
Finally, a variation of $\pm$0.2 km/s in $v_t$ leads to a change of $\pm$0.10-0.15 dex 
in [Fe/H] and smaller than 0.1 dex for the other abundance ratios.

\end{enumerate}

   \begin{figure}
   \centering
   \includegraphics[width=\columnwidth]{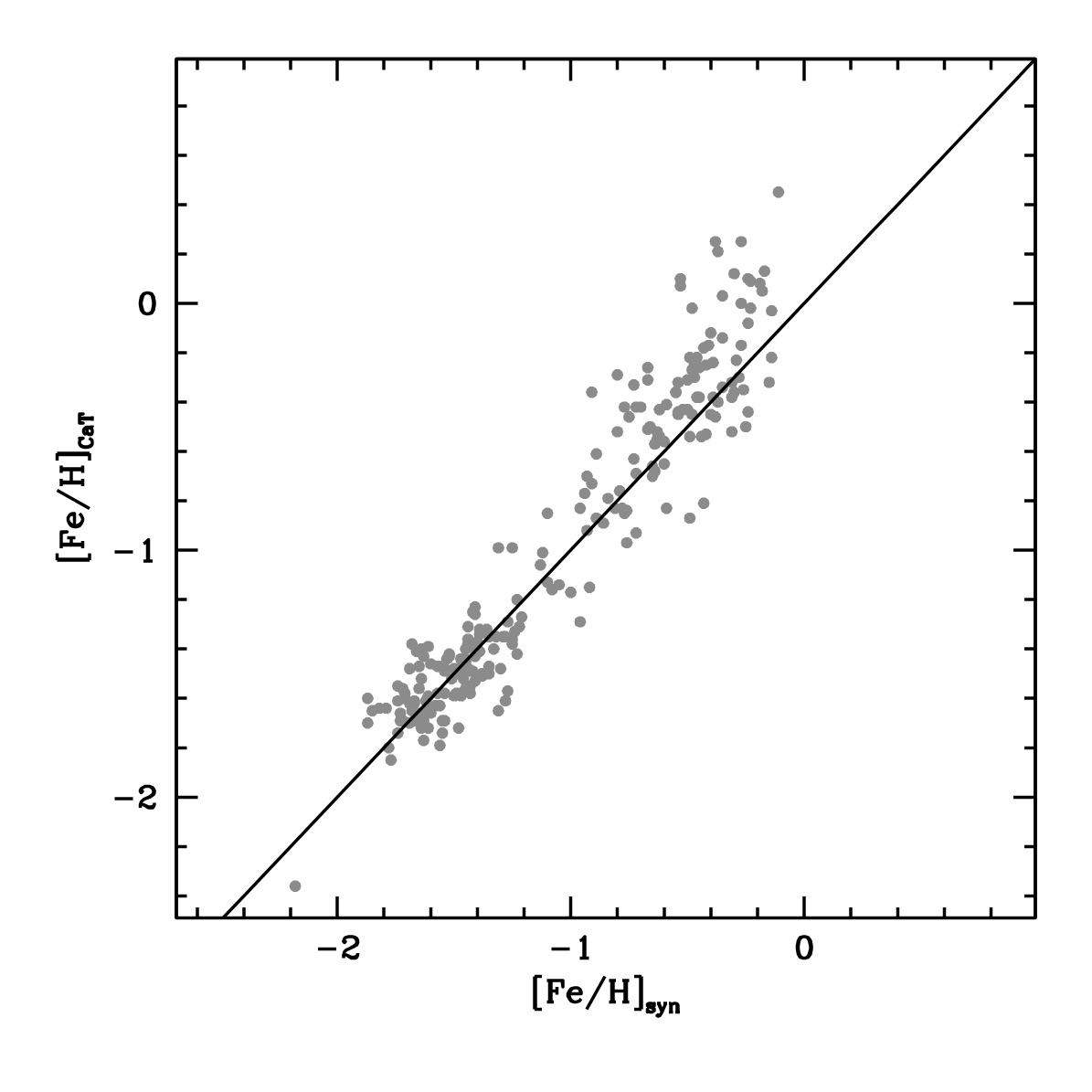}
     \caption{Comparison of iron abundances obtained by spectral synthesis of Iron lines in the present analysis and those
     derived by B08 with the Calcium triplet technique, from the same spectra. The diagonal line has slope=1.0.}
        \label{fecat}
    \end{figure}

   \begin{figure}
   \centering
   \includegraphics[width=\columnwidth]{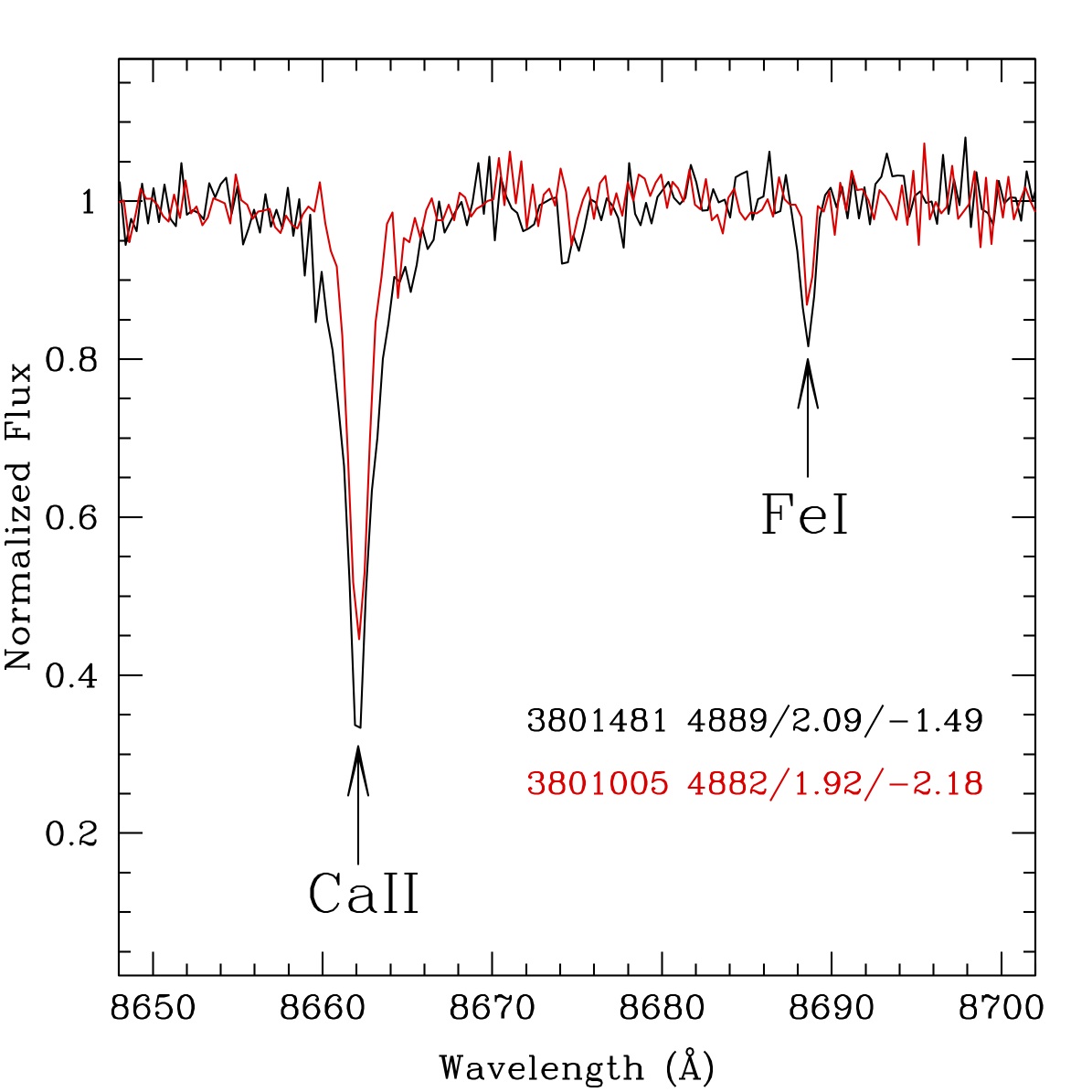}
     \caption{Comparison between the spectra of the stars \# 3801005 (red line) and 
     \# 3801481 (black line) around the Ca~II line at 8662 \AA\ and the Fe~I line at 8689 \AA .
     The two stars have similar atmospheric parameters 
     ($T_{eff}$=4882 K and log~g=1.92 for \# 3801005  and $T_{eff}$=~4889 K and log~g=2.09 for \# 3801481) 
     but different iron abundances.}
        \label{spec}
    \end{figure}


Finally, we derived abundances of Fe, Mg, Ca and Ti for 235 giant stars in M54 and Sgr,N.
This is by far the largest sample of chemical abundances in the main body of Sgr~dSph ever presented. 
While a large fraction of the sample is attributable to M~54 (i.e. the vast majority of the stars with -1.8$\la$[Fe/H]$\la$ -1.2), there are also 109 stars having [Fe/H]$\ge -1.0$  (hence unambiguously associated with Sgr,N), 
89 of which were never analysed before in terms of chemical composition. Hence this study more than doubles the number 
of stars with elemental abundances in the main metal-rich stellar component of Sgr~dSph.
In Fig.~\ref{cmd} the targets are plotted in the CMD according to their metallicity, red circles for 
the stars with [Fe/H]$\ge$-1.0 dex (associated to Sgr,N) and blue circles for those with [Fe/H]$<$-1.0 
(mainly stars belonging to M54). This figure is an analog to Figure 10 in B08, confirming that the metal-rich stars 
are all associated to the reddest RGB visible in the CMD and corresponding to Sgr,N. It is important to note 
that the metal-rich sample is largely composed of Red Clump (RC) stars, while the metal-poor one is made only of RGB stars.

\section{Comparison with previous studies}
\label{compa}

In Fig.~\ref{fecat} we show that the new [Fe/H] estimates based on synthetic spectral analysis are in agreement with those 
by B08 from the Calcium Triplet, within the uncertainties. A small systematic deviation is apparent in the most 
metal-rich regime where it is known that the CaT technique may loose some sensitivity and reliability
\citep[see e.g. the discussion in][]{carrera07}.

We note that one of the four Sgr,N members found by B08 to have [Fe/H]$_{\rm CaT}\le -2.0$ is included in the present sample 
(namely \# 3801005)
and is confirmed to have an iron abundance less than $\frac{1}{100}$ solar by spectral analysis of iron lines ([Fe/H]=--2.18$\pm 0.20$). 
Fig.~\ref{spec} shows the comparison between the spectrum of the star \#~3801005 and the star 
\# 3801481, with similar atmospheric parameters and a metallicity of [Fe/H]=--1.58 dex, compatible with the 
bulk of the M54 metallicity distribution. The difference in the metallicity of the two targets is clearly visible in the strengths 
of both the Fe~I and  Ca~II lines shown in Fig.~\ref{spec}. 
The good match with the CaT metallicity of B08 suggests that also the other four candidates found in that study are 
likely genuine very metal-poor stars. 
We stress that very few stars with metallicity [Fe/H]$\le-1.5$, as derived from direct spectroscopic measures, 
are known in the main body of Sgr \citep[see, e.g, MWM13 and Fig.~4 in][]{sbordone15}.

   \begin{figure}
   \centering
   \includegraphics[width=\columnwidth]{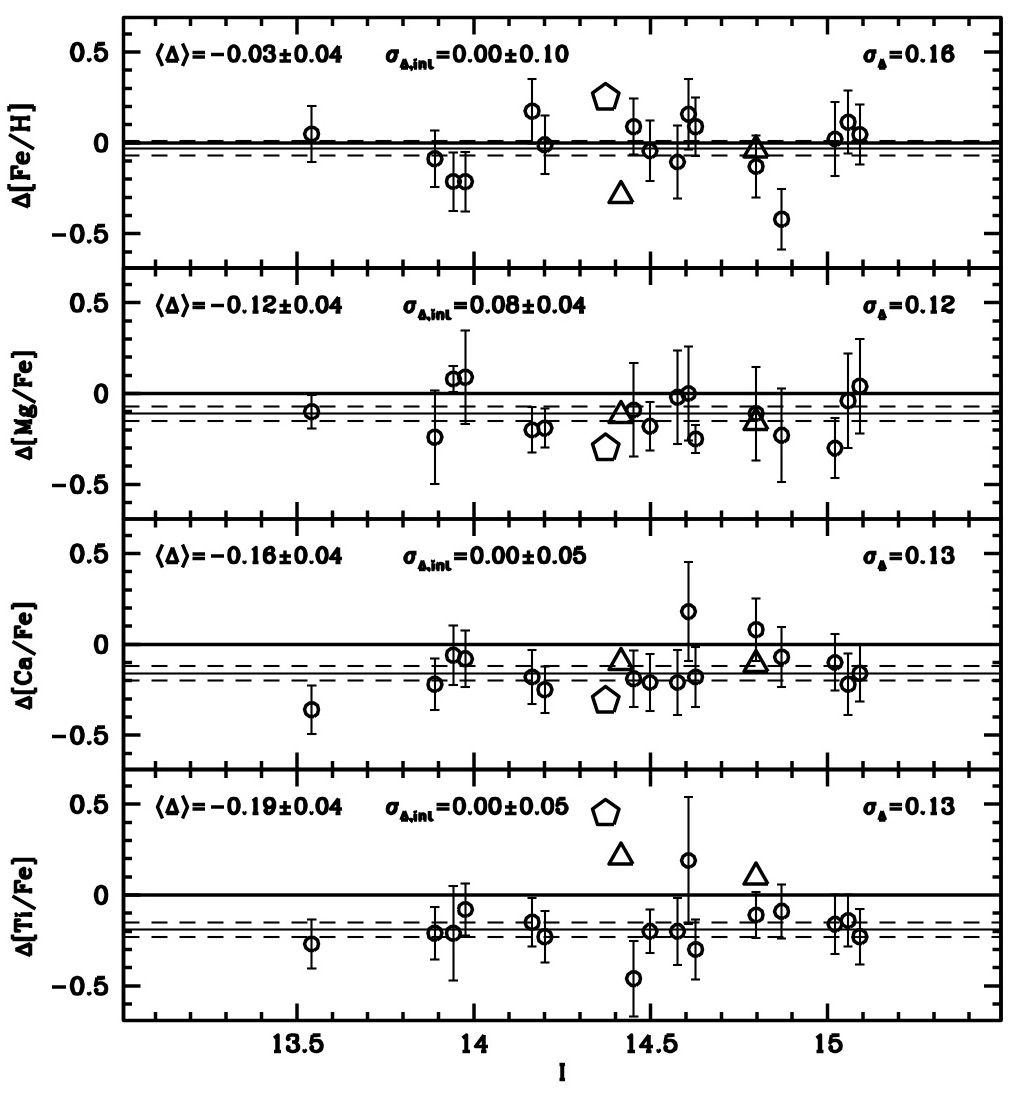}
     \caption{Differences in the abundance of stars in common between this work and other analyses ( $\Delta$= other analyses - 
     this work). The small empty circles are from C10b, empty triangles from M05, empty pentagons from MWM13. The thin 
     continuous line is the mean of the difference w.r.t. C10b and the thin dashed lines enclose the errors about the mean, 
     as derived with the ML algorithm (value reported in the upper left corner of each panel, together with the intrinsic 
     dispersion and the associated uncertainties). The value of the straight standard deviation of the difference 
     are also reported in the upper right corner of each panel.}
        \label{comuni}
    \end{figure}


In Fig.~\ref{comuni} we plot the difference in abundance ratios between our estimates and those previously 
available in the literature for stars in common. It is pretty clear that a fully meaningful comparison can be 
performed only with the C10b sample, with which we have sixteen stars in common (only one with [Fe/H]$>-1.0$, however), 
while two are in common with M05 and one in common with MWM13\footnote{For abundances derived from one single line C10b 
does not provide uncertainties. We exclude these values from this and following comparisons. The same is true for one 
of the stars in the sample by \citet{chou}.}. For this reason, and since the C10b analysis is based on higher resolution 
spectra we take it as a reference for comparison. The overall comparison is satisfactory.

We compute the mean and {\em intrinsic} standard deviation ($\sigma_{int}$) of the differences between the abundance 
ratios by C10b and our own ones using the Maximum Likelihood (ML) algorithm described in \citet{Mu12}. In all the cases 
$\sigma_{int}$ is consistent with zero, within the uncertainties. Indeed, the overall standard deviation reported in the 
upper right corner of each panel is always of the order of the uncertainties on individual abundance estimates.

The mean differences in the abundance ratios, with the associated uncertainties, are reported in the upper left corner of each panel of Fig.~\ref{comuni}.
While the mean difference in [Fe/H] is null, within the uncertainties, there are small but significant differences in 
the zero point of abundance ratios involving the $\alpha$-elements. We note that for [Mg/Fe] and [Ca/Fe] the three 
points from M05 and MWM13 are consistent with the distribution of C10b points. This is not the case for [Ti/Fe]; 
we will see below that this ratio is indeed more problematic (see also MWM13 for a discussion). To be compliant with C10b 
we apply to all our abundance ratios the mean shifts reported in Fig.~\ref{comuni}. The estimates used in the following 
and listed in Table~\ref{abu} have been corrected accordingly.

\begin{table*}
  \begin{center}
  \caption{Atmospheric parameters and abundances [the full table is available in the electronic version].}
  \label{abu}
  \begin{tabular}{lccccccccccccc}
     ID & RA$_{J2000}$&Dec$_{J2000}$ &T$_{eff}$&log g& v$_t$&[Fe/H]& err &[Mg/Fe]& err  &[Ca/Fe]& err  &[Ti/Fe]& err \\
        &  [deg]      & [deg]	     &[$^o$K] &{\tiny [cm s$^{-2}$]}& {\tiny [km s$^{-1}]$} &	 &	 &	&	&      &       &      &  \\
\hline
3700223& 283.655137 & -30.398678  & 4581 & 1.47 & 1.8 & -1.47 & 0.14 &  0.40 & 0.08 &  0.33 & 0.12 &  0.08&  0.08\\
3700626& 283.657212 & -30.400988  & 4804 & 2.02 & 1.7 & -1.44 & 0.16 &  0.39 & 0.12 &  0.44 & 0.13 &  0.25&  0.11\\
3800223& 283.813879 & -30.403892  & 4060 & 0.60 & 2.0 & -1.21 & 0.14 &  0.09 & 0.07 &  0.29 & 0.11 &  0.07&  0.09\\
3800306& 283.819591 & -30.464056  & 4586 & 1.32 & 1.8 & -1.28 & 0.14 &  0.44 & 0.07 &  0.20 & 0.11 & -0.05&  0.12\\
3800422& 283.781041 & -30.463381  & 4460 & 1.38 & 1.8 & -1.39 & 0.14 &  0.25 & 0.08 &  0.36 & 0.11 &  0.16&  0.11\\
3800471& 283.718195 & -30.441672  & 4498 & 1.36 & 1.8 & -1.57 & 0.15 &  0.41 & 0.08 &  0.33 & 0.13 &  0.06&  0.09\\
3800508& 283.725104 & -30.440299  & 4614 & 1.51 & 1.8 & -1.68 & 0.15 &  0.38 & 0.08 &  0.38 & 0.11 &  0.07&  0.08\\
... &&&&&&&&&&&&&\\
\hline
\end{tabular} 
\end{center}
\end{table*}

\begin{figure}
   \centering
   \includegraphics[width=\columnwidth]{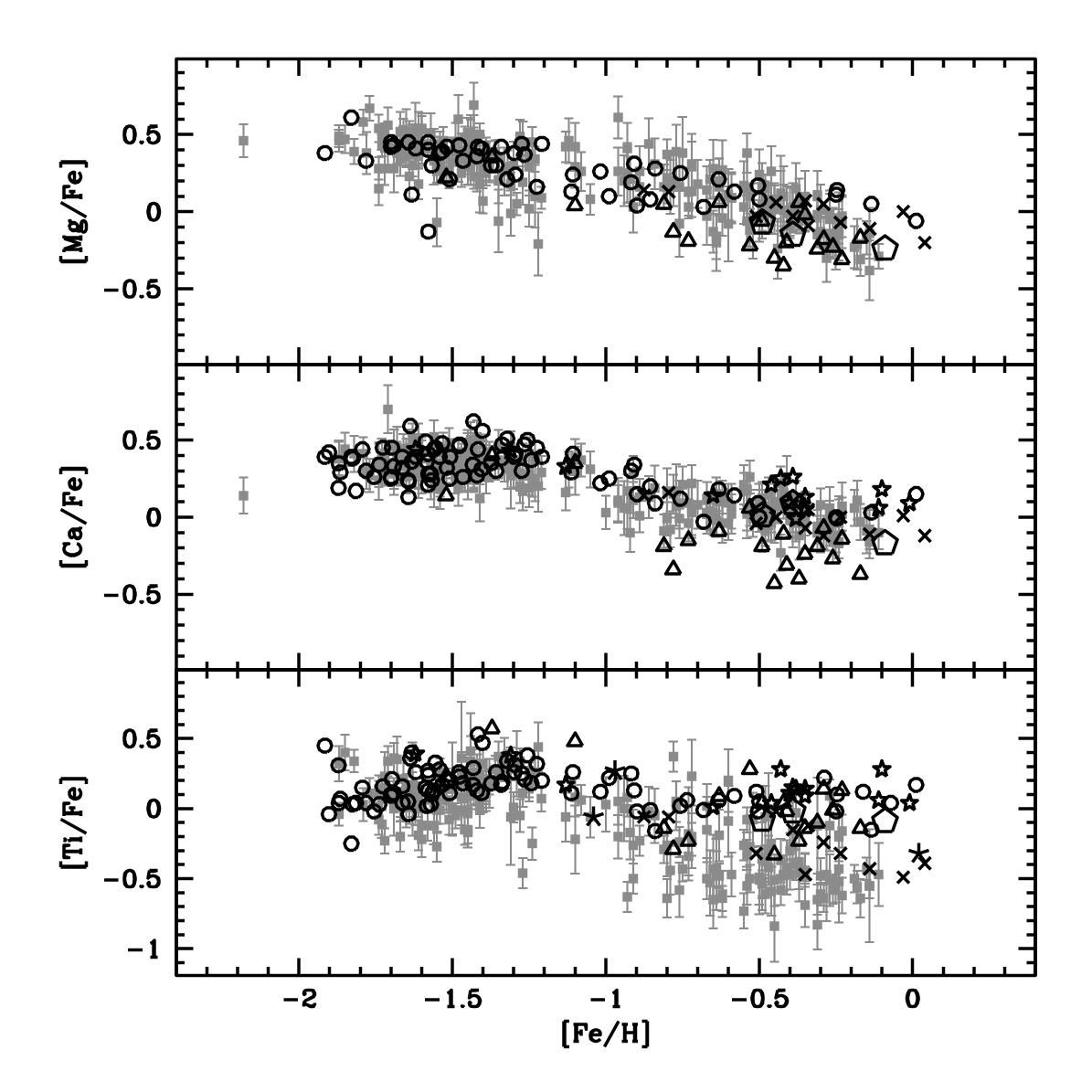}
     \caption{Comparison of the trends of the $\alpha$-elements considered here vs. [Fe/H] obtained from the 
     present analysis (small grey squares) with datasets from the literature. Small empty circles: C10b; 
     $\times$ symbols: S07; empty stars: SM02; empty triangles: M05; empty pentagons: MWM13; asteriks: \citet{chou}.}
        \label{alldat}
    \end{figure}

In Fig.~\ref{alldat} we compare our results with all the abundances for stars in the main body of Sgr available 
in the literature in the planes opposing [Mg/Fe], [Ca/Fe], and [Ti/Fe] to [Fe/H]. The overall agreement is good 
for [Mg/Fe] and [Ca/Fe], especially if the variety of sources is considered. The agreement is satisfactory also 
for [Ti/Fe] up  to [Fe/H]$\simeq -0.8$. However at larger metallicity the spread between different sets of measures 
is large and probably dominated by systematics. In particular, the bulk of our estimates is significantly lower 
than the [Ti/Fe] range spanned by most of literature values. 
However, we note that a significant scatter does exist also among the different [Ti/Fe] trends with [Fe/H]
found by other authors, in the metal-rich stars of Sgr,N. The origin of the discrepancy is not clear: the Ti lines are 
very sensitive to temperature and microturbulent velocity and we cannot rule out that systematic uncertainties 
in the $T_{eff}$ scale for the metal-rich stars may lead to systematic mis-estimates of the Ti abundance (see also MWM13).
Due to these uncertainties in the analysis and to the uncertainties in the Ti yields that affect the chemical evolution models (see Sect.~\ref{chem_mod}), we exclude this element from the following discussion.

From the comparisons above we conclude that our Fe, Mg and Ca abundances are reliable over the whole range of 
metallicity spanned by our data, while [Ti/Fe] values in the metal-rich regime must be considered with great caution.
In the following we will limit our analysis to our own sample to take advantage of the highest degree of internal 
homogeneity and consistency of our measures. This is an important step forward in our view of the chemical patterns 
in Sgr~dSph, since the mutual consistency of the abundance sets from various authors is not well established; 
systematic differences have been noted (see, e.g., MWM13) and can be perceived also in Fig.~\ref{alldat}.

\section{Chemical patterns in Sgr,N}
\label{chem}

   \begin{figure}
   \centering
   \includegraphics[width=\columnwidth]{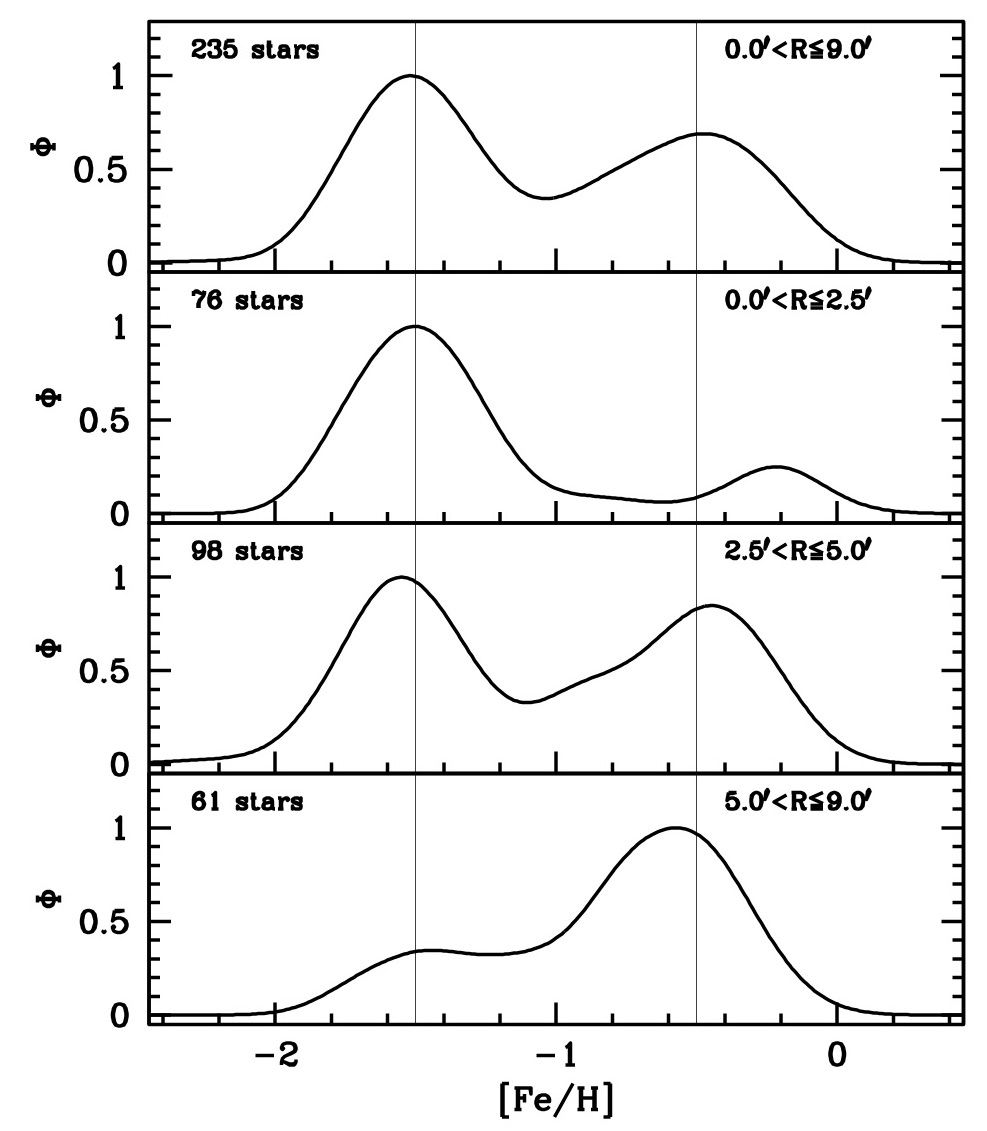}
     \caption{Metallicity distribution of stars in our sample in different radial annuli around 
     the center of M~54/Sgr,N. The distributions are in the form of generalised histograms normalised at their maximum. 
     The thin vertical lines are for reference and are placed at [Fe/H]=-1.50 and [Fe/H]=-0.5, near the mean of the two peaks, corresponding to M~54 and Sgr,N. }
        \label{distFe1}
    \end{figure}


   \begin{figure}
   \centering
   \includegraphics[width=\columnwidth]{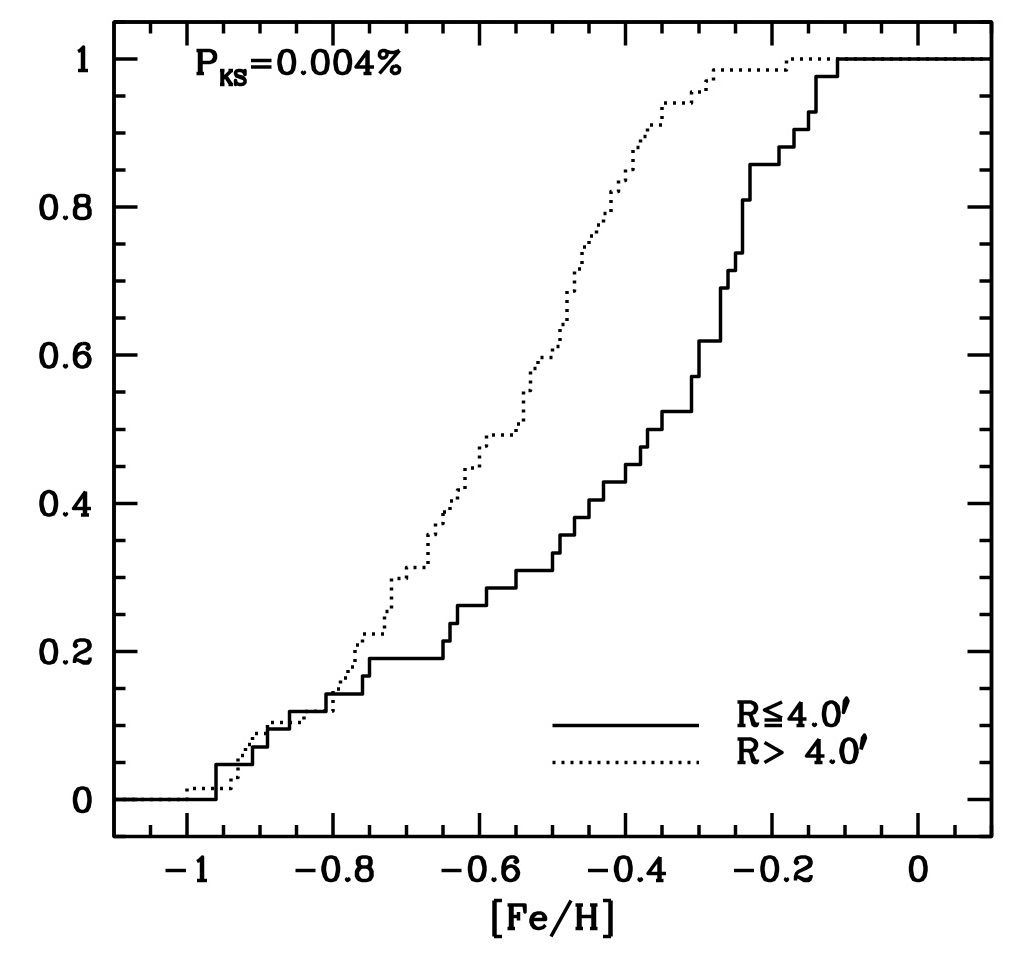}
     \caption{Cumulative metallicity distribution for stars with [Fe/H]$\ge -1.0$ 
     in two different radial ranges. $P_{KS}$ is the probability that the two distributions be extracted from the same 
     parent population according to a Kolmogorov-Smirnov test.}
        \label{distFe2}
    \end{figure}


   \begin{figure}
   \centering
   \includegraphics[width=\columnwidth]{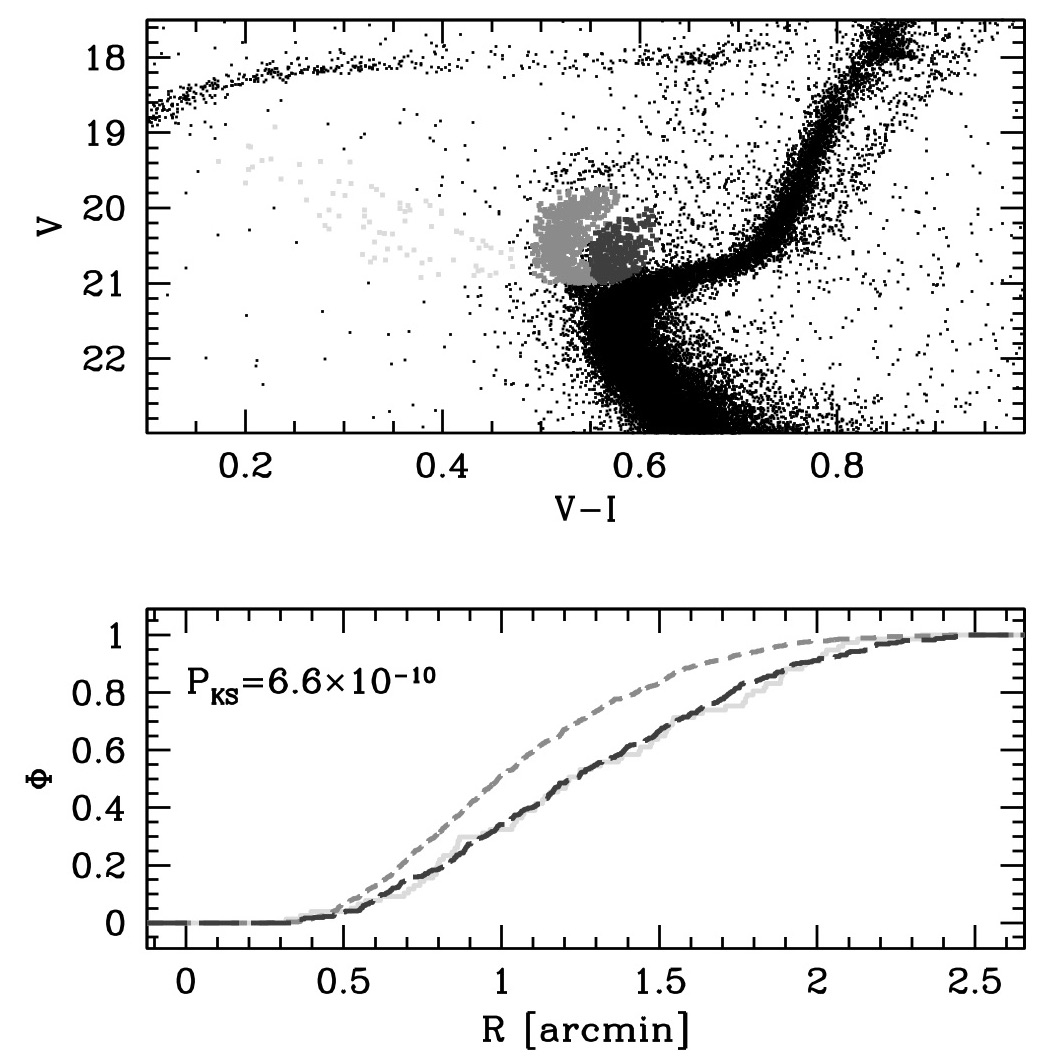}
     \caption{Upper panel: color magnitude diagram of the innermost $2.5\arcmin$ from the center of Sgr,N from the ACS-HST photometry of \citet{siegel}. The stars selected in 
     the CMD features associated by these authors with a possible young population (or a population of BSS), a population with age$\simeq 2.3$~Gyr and [Fe/H]=-0.1, 
     and a population with age$\simeq 4-6$~Gyr and [Fe/H]=-0.6, are highlighted in different tone of grey, from the palest to the darkest, respectively. 
     Lower panel: the cumulative radial distribution of the three populations are plotted using the same tones of grey (moreover, continuous line: young/BSS pop., 
     short-dashed line: 2.3~Gyr pop., long-dashed line: 4-6 Gyr population).
The probability that the samples tracing the 2.3~Gyr and the 4-6~Gyr populations have the same radial distribution is reported.}
        \label{hst_rad}
    \end{figure}


In  Fig.~\ref{distFe1} we show the metallicity distribution of our sample in different radial annuli. 
We recall again that this distribution is biased by the inclusion of M~54 {\em and} by the strong selection toward stars 
lying near the main RGB/RC sequences in the CMD, i.e. toward stars belonging to M~54 or to the main metal-rich population of Sgr.
Indeed these populations are well traced by the two strong peaks that characterise the total distribution in the 
upper panel of Fig.~\ref{distFe1}. Still, moving to more external annuli, with the progressive decreasing of the M~54 peak, 
the metal poor tail of the Sgr,N population becomes discernible as a left wing asymmetry of the metal-rich peak.

Using the ML algorithm on various subsamples we tried to estimate the mean metallicity of the populations corresponding 
to the main peaks and to put constraints on their intrinsic dispersion. Attempting to isolate the purest sample of 
bona-fide M~54 members we selected 61 stars having $-2.0<$[Fe/H]$<-1.0$ and $R\le 2.5\arcmin$. From this subsample we 
obtain $\langle {\rm [Fe/H]}\rangle = -1.52\pm 0.02$ and $\sigma_{int}=0.09\pm 0.03$, in good agreement with C10b. For the overall sample
of the Sgr,N metal-rich component, i.e. the 109 stars with [Fe/H]$\ge -1.0$, we find $\langle {\rm [Fe/H]}\rangle = -0.52
\pm 0.02$ and $\sigma_{int}=0.17\pm 0.02$. Even with the cut at [Fe/H]=--1.0, the metallicity distribution of Sgr,N is much 
broader than that of M~54.
We note that the modes of the distribution shown in the upper panel of Fig.~\ref{distFe1} occur 
at [Fe/H]=--1.52 and [Fe/H]=--0.47 dex, hence the ML means above traces pretty well the position of the two 
metallicity peaks.

In the last years several studies addressed the metallicity distribution of stars in the Sagittarius stream, with 
different techniques. All of them found a metallicity gradient along the Stream 
\citep[see, e.g.,][and references therein]{b06a,mo07,chou,keller,hyde,gibbons} hinting at a pre-existing metallicity gradient 
in the population of the progenitor.
Indeed, photometric evidences of a metallicity gradient within the main body were already presented by \citet{sdgs2} 
and \citet{alard}. Here we present the first spectroscopic evidence for a metallicity gradient in the main body of Sgr~dSph, 
since Fig.~\ref{distFe1} {\em shows that the peak of the metal-rich population changes with projected distance from the 
center within the stellar nucleus of Sgr}. This peak is at [Fe/H]$=-0.38$ for $0.0\arcmin<R\le 2.5\arcmin$, at 
[Fe/H]$=-0.45$ for $2.5\arcmin<R\le 5.0\arcmin$, and at [Fe/H]$=-0.57$ for $5.0\arcmin<R\le 9.0\arcmin$. As weak as it may appear, 
the change of metallicity distribution of the metal-rich ([Fe/H]$\ge -1.0$) component is demonstrated to be real and 
statistically significant in Fig.~\ref{distFe2}. The probability that the Sgr,N stars within and 
outside a radius of 4.0$\arcmin$ from the center are extracted from the same parent distribution of [Fe/H] is $P=0.004$ per cent, 
according to a Kolmogorov-Smirnov test, the innermost sub-sample being clearly skewed toward higher metallicities. It is 
unclear if the gradient on such a small scale ($\la 100$~pc) is related to the kpc-scale gradient that is seen in the 
main-body at large and in the Stream. Still it provides a relevant additional constraint to models for the formation of the Sgr nucleus.

In their analysis of the stellar populations in the innermost $2.5\arcmin$ of Sgr,N, based of ACS-HST data, \citet{siegel} associated the multiple turn-offs (TO) seen in their CMD (above the old and metal-poor one dominated by M54 stars) with up to four subsequent bursts of star formation, with increasing metallicity. It is tantalising to conclude that the metallicity gradient described above corresponds to an age gradient, with more recent and more metal rich bursts occurring 
closer to the center of the system, i.e. with a smaller characteristic length scale. In particular \citet{siegel}
identifies a population with age=4-6~Gyr and [Fe/H]$\simeq -0.6$ (possibly split into-two sub populations), a
population with age=2.3~Gyr and [Fe/H]$\simeq -0.1$, and a blue plume population that can be interpreted as
having age=0.1-0.8~Gyr and [Fe/H]$\simeq +0.6$ or as being largely composed by Blue Stragglers \citep[BSS, see][and references therein]{f16}.
This multi-modality in mean age/metallicity  is not reflected in the metallicity distribution shown in 
Fig.~\ref{distFe1}, but this does not seem particularly significant since multiple peaks in a relatively narrow metallicity range would be easily 
wiped out by the effect of observational uncertainty on individual metallicity estimates.

On the other hand, we used the \citet{siegel} 
photometry\footnote{Retrieved from {\tt \tiny http://www.astro.ufl.edu/\~ata/public\_hstgc/}} (adopting similar selections on quality parameters) to compare the radial distribution of the three main sub-populations. The adopted selection strictly follows the age/metallicity classification by \citet{siegel} and is illustrated in the upper panel of Fig.~\ref{hst_rad}. The TO stars used to trace the various populations can be clearly recognised in the CMD, they are more than six magnitude above the limiting magnitude of the overall photometry, and cover similar ranges in magnitude, therefore they should have very similar completeness as a function of distance from the center of the system. For this reason, their radial distributions can be safely compared, as done in the lower panel of Fig.~\ref{hst_rad}. This plot shows that the age=2.3 Gyr/[Fe/H]$\simeq -0.1$ population is more centrally concentrated than the age=4-6 Gyr/[Fe/H]$\simeq -0.6$, for $R\le 2.5\arcmin$. This implies that the more metal-rich population dominates the innermost region while the one with mean metallicity [Fe/H]$\simeq -0.6$ is likely to become more and more relevant at larger radii, in excellent agreement with our findings based on abundance analysis. According to a Kolmogorov-Smirnov test, the probability that the radial distribution of the two samples is extracted from the same parent distribution is $P_{KS}=6.6\times 10^{-10}$. The radial distribution of blue plume stars is indistinguishable from that of the older population, suggesting that it is most likely largely composed of BSS from both M54 and Sgr,N. 

Hence both photometric and spectroscopic evidence independently suggest that the metal-rich population of Sgr,N formed outside-in over a timescale of a few Gyr while the mean metallicity grew from $\simeq 1/4$ solar to solar.

   \begin{figure}
   \centering
   \includegraphics[width=\columnwidth]{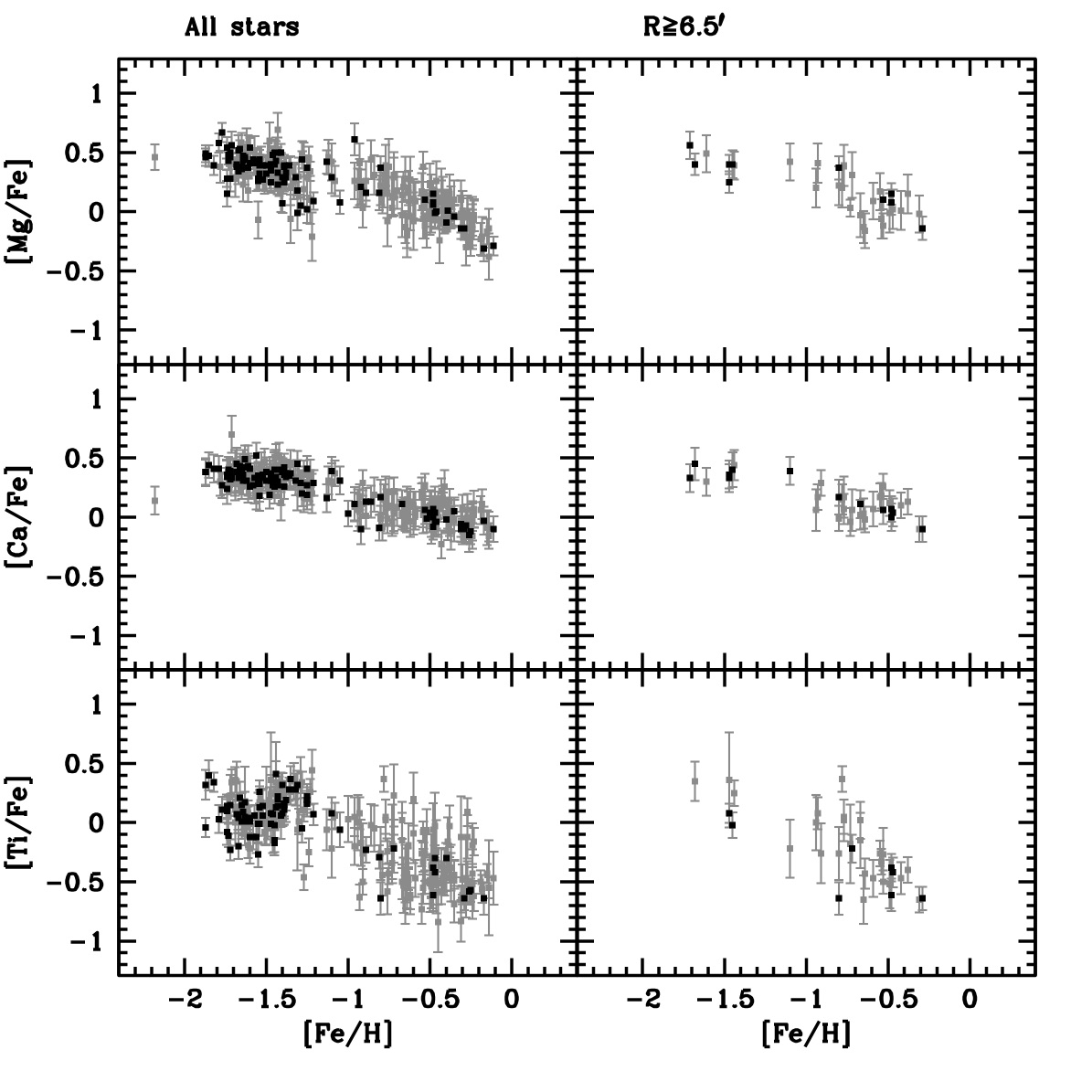}
     \caption{Trend of all the elemental abundances considered here vs. [Fe/H] for the whole sample (left panels) and 
     for a sub-sample minimising the contribution from M54~stars (right panel; R$\ge 6.5\arcmin$). The black dots correspond 
     to elemental ratios whose uncertainty is $\le 0.15$~dex.}
        \label{alfa}
    \end{figure}


\subsection{Trends with metallicity}

In the left column of panels of Fig.~\ref{alfa} we show the trends of [Mg/Fe], [Ca/Fe] and [Ti/Fe] against iron abundance 
for the entire sample. The comparison of the total sample with the subset of targets with uncertainties on the abundance 
ratio estimates lower than 0.15~dex (black points) suggests that a significant part of the spread seen at each metallicity 
is due to observational errors and the underlying {\em true} trends are relatively narrow in the [$\alpha$/Fe] direction.

In the right column of panels we make an attempt to minimise the contamination by M~54 stars by selecting only the handful 
of stars with $R\ge 6.5$, the radius out of which the surface brightness of the cluster drops below that of Sgr, 
according to B08 (their Fig.~2, in particular). The overall trends remain similar to those including the full M54~population. The data  available in the literature suggest that the mean $\alpha$-elements abundance of M~54 
is consistent with that of Sgr~dSph as a whole in the relevant range of metallicity. 
The analysis of the Keck/DEIMOS spectra supports this view, consequently we will assume, as a working hypothesis, that M~54 stars are good 
tracers for the  $\alpha$-elements abundance trends  of Sgr in the metal-poor regime ({\em but not of the radial distributions}, as explained
above).

\begin{figure}
   \centering
   \includegraphics[width=\columnwidth]{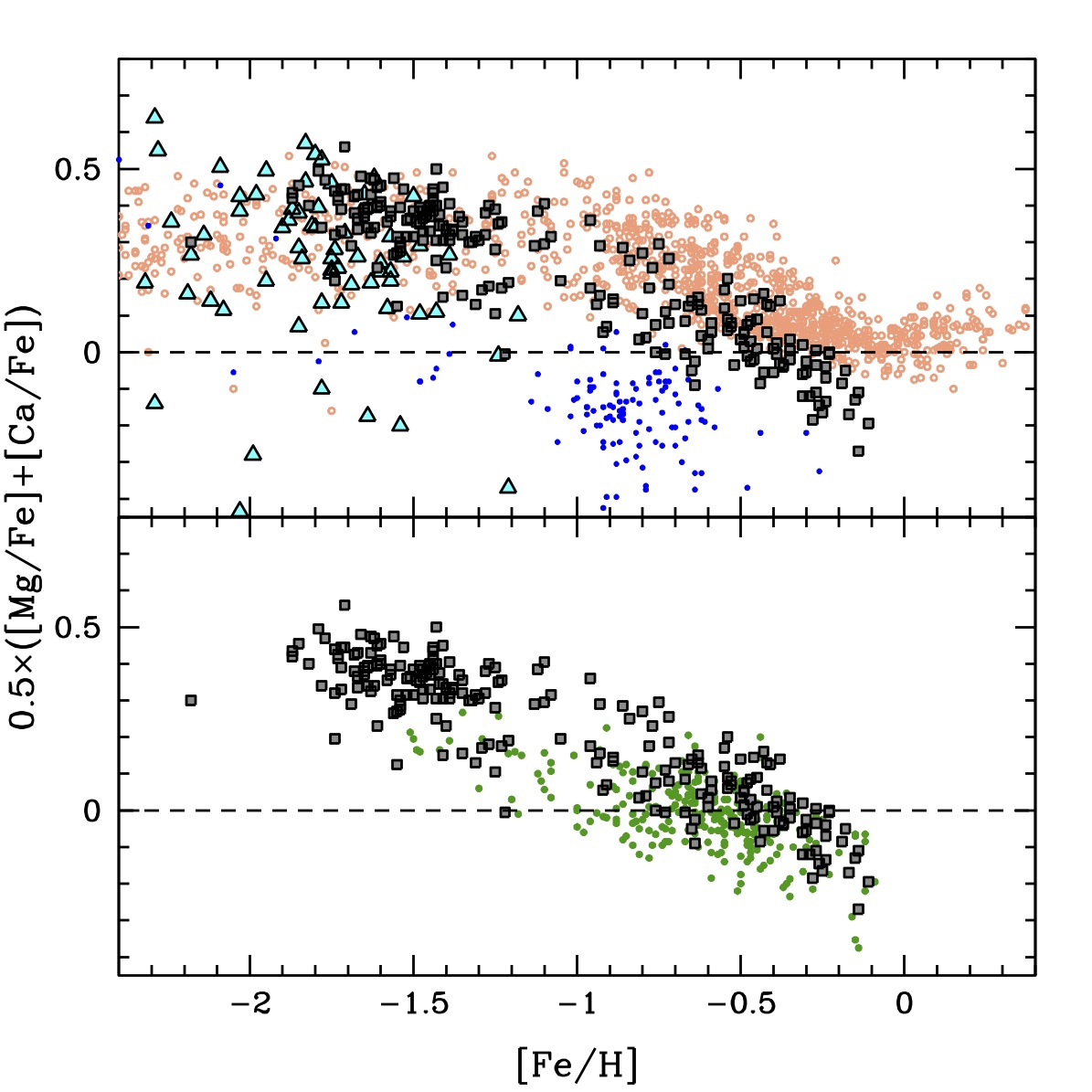}
     \caption{Behaviour of $0.5\times$([Mg/Fe]+[Ca/Fe]) as a function of [Fe/H] 
     for the whole sample (grey squares) in comparison with (upper panel) the Milky Way stars 
     \citep[red circles,][]{edvardsson93, burris00, fulbright00, stephens02,   
     gratton03, reddy03, barklem05, bensby05, reddy06}, Fornax \citep[blue dots,][]{letarte10,lemasle14} and 
     Carina stars \citep[cyan triangles,][]{lemasle12,fabrizio15}, 
     and with (lower panel) LMC stars \citep[green dots,][]{lapenna12,swaelmen13}.}
        \label{alfa2}
    \end{figure}

Fig.~\ref{alfa2} compares the [$\alpha$/Fe] abundance ratio  
(defined as $0.5\times$([Mg/Fe]+[Ca/Fe])) as a function of [Fe/H] of our Sgr,N sample with samples from different galaxies taken from
the literature. The galaxies we compare with are:
the Milky Way \citep[red circles,][]{edvardsson93, burris00, fulbright00, stephens02,   
gratton03, reddy03, barklem05, bensby05, reddy06}, 
Fornax \citep[blue dots,][]{letarte10,lemasle14} and 
Carina \citep[cyan triangles,][]{lemasle12,fabrizio15}, in the upper panel,
and LMC  \citep[green dots,][]{lapenna12,swaelmen13}, in the lower panel of Fig.~\ref{alfa2}.

The metal-poor component of our sample ([Fe/H]$<-1.0$, dominated by M54 stars) exhibits an enhancement of the 
[$\alpha$/Fe] abundance ratio that matches well with those of the Galactic Halo stars 
and the bulk of the Carina stars.  On the other hand, in the metal-rich regime the level of $\alpha$-elements
abundance is intermediate between the MW (on the high side) and Fornax (on the low side), at any metallicity.
The inspection of the lower panel of Fig.~\ref{alfa2} reveals that the overall pattern of $\alpha$-elements vs. [Fe/H] seen in the 
Sgr~dSph is quite similar to that observed in the LMC, suggesting a similarity between the progenitor of Sgr~dSph 
and the most massive of the satellites that orbit the Milky Way \citep[see also][]{mo05,deboer,gibbons}. 

\begin{figure}
   \centering
   \includegraphics[width=\columnwidth]{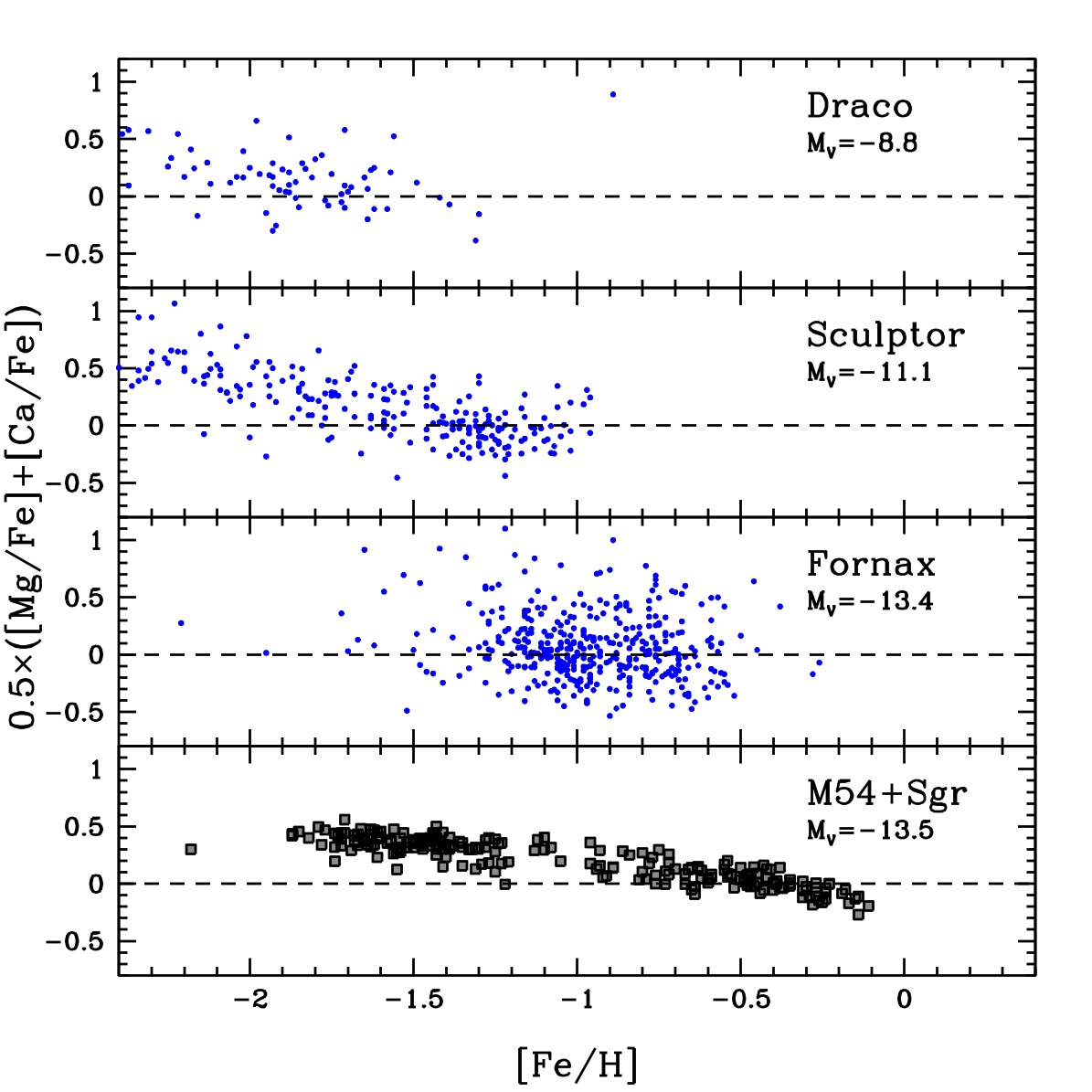}
     \caption{Behaviour of $0.5\times$([Mg/Fe]+[Ca/Fe]) as a function of [Fe/H] 
     for the whole sample (grey squares, lowest panel) in comparison with 
     the stars of Draco, Sculptor and Fornax dSph's analysed by \citep[blue points,][upper panels]{kirby10} 
     using Keck/DEIMOS spectra with the same set-up as those analysed here.}
        \label{alfa3}
    \end{figure}

A high mass progenitor is hinted at also by the comparison with other dwarf spheroidal galaxies shown in Fig.~\ref{alfa3}. 
The $0.5\times$([Mg/Fe]+[Ca/Fe]) vs. [Fe/H] patterns of Draco, Sculptor and Fornax dSph's are taken from \citet{kirby10}, 
who measured their abundance estimates from Keck/DEIMOS spectra with the same set-up as those analysed here. In this figure, 
the galaxies are ordered in the various panels according to their V-band 
absolute magnitude, a proxy for stellar mass. A trend of the maximum metallicity with $M_V$ is clearly evident, suggesting 
that the progenitor of Sgr dSph was more massive than Fornax dSph, that is the most massive among dSphs orbiting the Milky Way.

To further investigate the chemical evolution of the Sgr system and the properties of its progenitor/precursor, 
we recur to chemical evolution models.

\subsection{Chemical evolution of the Sagittarius complex precursor}
\label{chem_mod}

Chemical evolution models for Sgr in the literature refer to its bound central 
region \citep[e.g.][]{lanmat03,vin15}; in this section, we discuss a more 
general model for a precursor that interacts disruptively with the Galaxy 
leaving behind the compact body and the stellar streams we see today. It is evident
that both theoretical limitations and the lack of robust observational constraints (e.g.
on the total mass of the progenitor, the epoch of infall, etc.) hampers our
ability to model in proper detail a system with a complex chemo-dynamical evolution like the Sgr~dSph.
Here we use a pure chemical model to reproduce the observed chemical abundance pattern 
to get broad indications on the past evolutionary path of the system.

\cite{nied10} have re-assembled the stellar debris of the Sgr dSph, and found 
that the stellar mass of the progenitor at infall must have been as high as 
$M_\star \simeq$~6.4\,$\times$10$^8$~M$_\odot$ \citep[see also][]{nied12}, i.e. 
more than a factor of 10 higher than the present-day stellar mass of the main 
body. From this, and using the stellar mass-halo mass relation from 
\cite{beh10}, it turns out that the progenitor of Sgr was embedded in a heavy 
dark halo with $M_{\mathrm{DM}} \sim$ 10$^{11}$~M$_\odot$. In order to reproduce 
the run of velocity dispersion with longitude for the metal-poor and metal-rich 
populations in the Stream, \cite{gibbons} set a lower limit of 
6\,$\times$10$^{10}$ M$_\odot$ to the dark halo mass. 

Resting on the above-mentioned observational evidence, we follow the chemical 
evolution of a system with $M_{\mathrm{DM}}$~= 6\,$\times$10$^{10}$~M$_\odot$ and 
$M_{\mathrm{bar}}$~= 0.17\,$M_{\mathrm{DM}}$~= 10$^{10}$~M$_\odot$. A fraction of 
$M_{\mathrm{bar}}$ is accreted in the form of cold neutral gas according to an 
exponentially-decreasing law, d$M_{\rm{inf}}$/d$t \propto \rm{e}^{-t/\tau}$, where 
$M_{\rm{inf}}$~= 2.1\,$\times$\,10$^9$~M$_\odot$ and $\tau$~= 0.5~Gyr. This gas is 
turned into stars following a Kennicutt-Schmidt \citep{sch59,ken98} law, 
$\psi(t) = \nu\,M_{\rm{gas}}(t)$, where $\nu$~= 0.1\,--\,0.001~Gyr$^{-1}$ is the 
star-formation efficiency and $M_{\rm{gas}}(t)$ is the mass of cold neutral gas 
at a given time. A more detailed description of the adopted formalism can be 
found in section~3.1 of \citet[][and references therein]{rom15}.

The star formation history of the precursor of Sgr as a whole is poorly-constrained; 
resting on studies by \cite{dol02} and \cite{deb15}, we assume that the bulk of 
the stars formed rather efficiently ($\nu$~= 0.1~Gyr$^{-1}$) from $\sim$14 to 
7~Gyr ago, then the star formation rate dropped abruptly ($\nu$~= 
0.001~Gyr$^{-1}$), until the star formation stopped 6~Gyr ago. This assumption may appear in stark
contrast with the results of \citep{siegel}. However this should be considered
just as a convenient working hypothesis at this stage. We will show below that 
we find good (possibly better) match with the observed chemical pattern also with
models in which the abrupt stop of star formation occur much later, while it is impossible
to reproduce the observations without a significant gas-loss episode.  
The model assumes a \cite{krou02} initial mass function (IMF), 
with a slope $x$~= 1.7 ($x$~= 1.35 for the Salpeter IMF) in the high-mass 
regime. Given the relatively deep potential well in which our Sgr precursor is 
embedded, the internal feedback alone is unable to trigger an outflow and 
deprive the galaxy from its residual gas. We thus assume that, starting from 
7.5~Gyr ago, because of the effects of Galactic tides the galaxy rapidly loses 
its gas, together with the outermost stellar populations that form the stream.
This gas stripping episode leads to the consumption of all the available gas
6~Gyr ago, in the model. Summarising the sequence of events, for clarity: (a) the model form stars
with $\nu$~= 0.1~Gyr$^{-1}$ since 14~Gyr ago; (b) since 7.5 Gyr ago gas is lost at a rate
tuned to reproduce the observed abundance patterns; (c) this gas loss process leads to the total consumption of the gas
reservoir by 6~Gyr ago, when, consequently, the star formation ceases.

\begin{figure*}
\begin{tabular}{ccc}
\includegraphics[width=5.5cm]{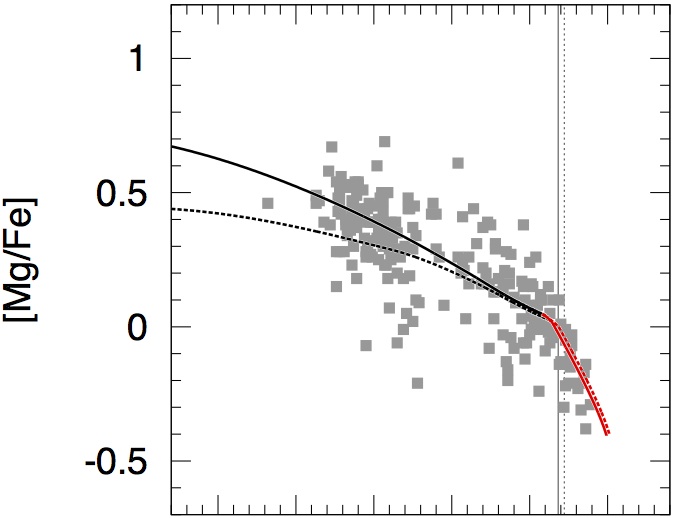} &
\includegraphics[width=5.5cm]{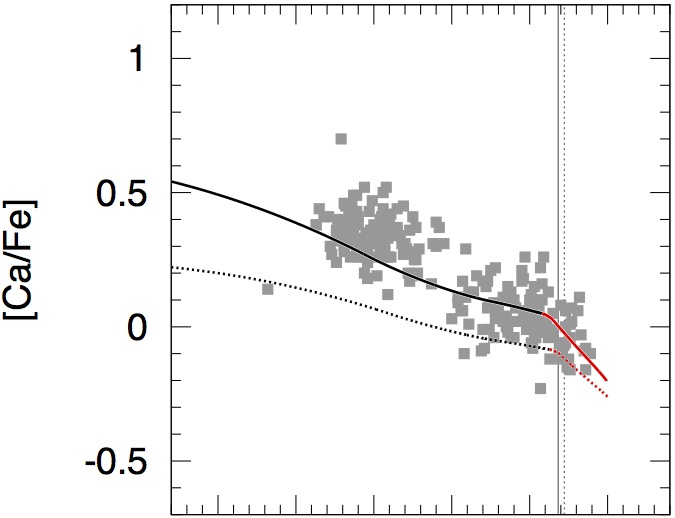} & 
\includegraphics[width=5.5cm]{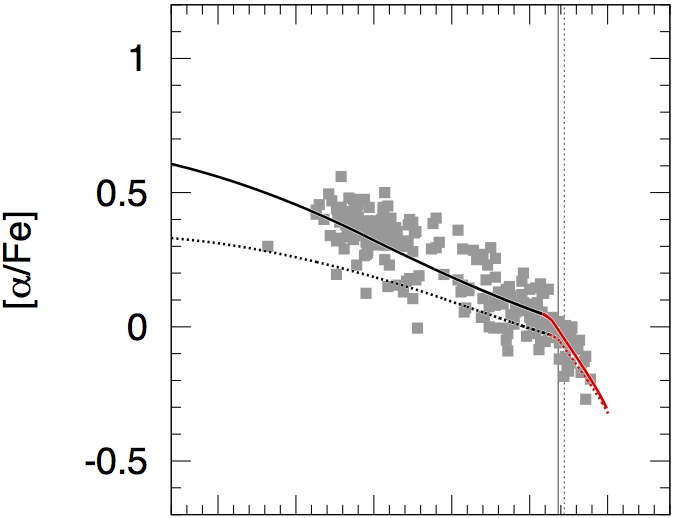} \\
\includegraphics[width=5.5cm]{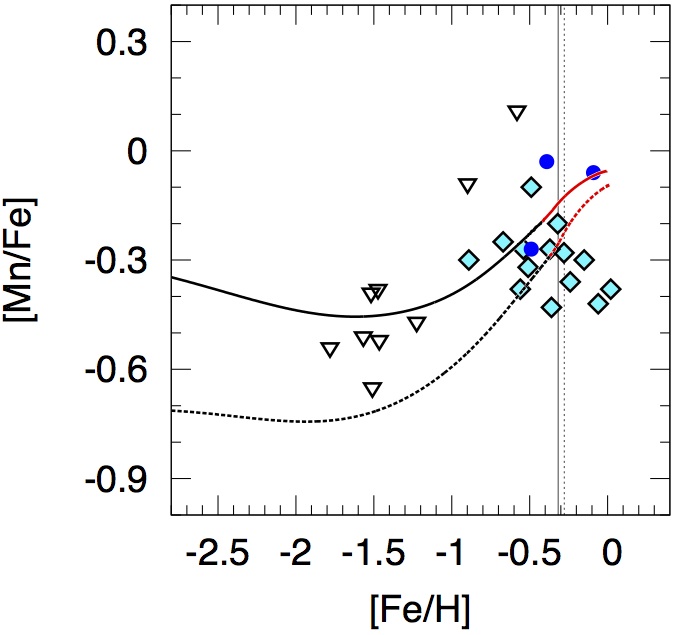} &
\includegraphics[width=5.5cm]{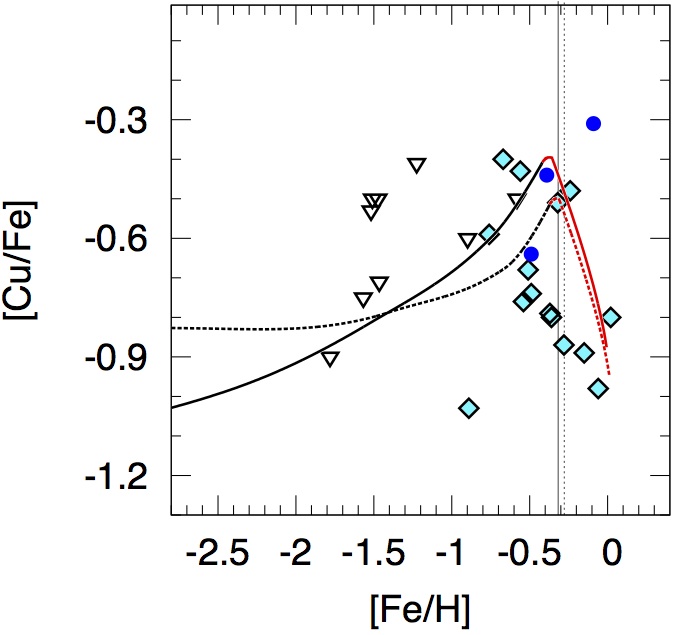} & 
\includegraphics[width=5.5cm]{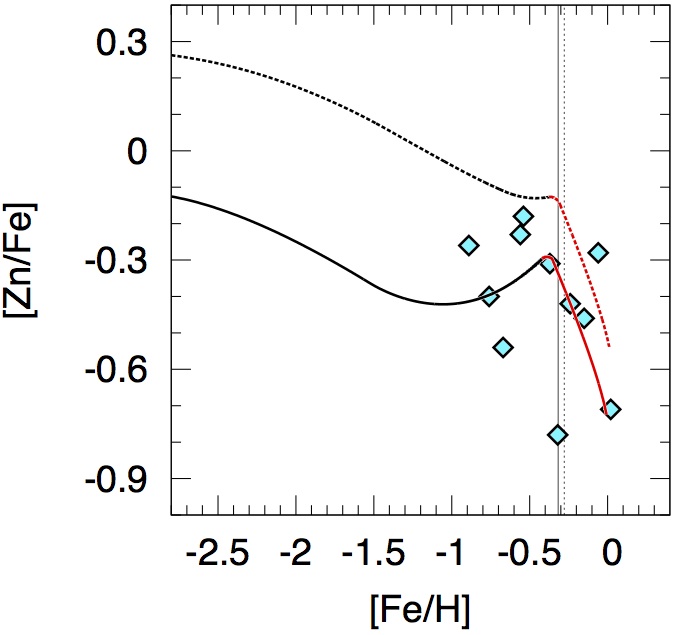} \\
\end{tabular}
\caption{Run of [X/Fe] versus [Fe/H] for --clockwise from top left-- Mg, Ca, 
$\alpha$-elements, Zn, Cu, and Mn predicted with our chemical evolution model 
assuming that stars above 20~M$_\odot$ explode as normal type II supernovae 
(solid curves) or as hypernovae (dotted curves). The red portions of the curves 
show the phases in which the stripping is active, while the vertical lines mark 
the endpoints of the tracks without stripping. The theoretical predictions are 
compared to data from this work (grey filled squares), 
\citet[][cyan diamonds]{sbo07}, 
\citet[][upside-down triangles]{c10b} and 
\citet[][blue filled circles]{mcw13}.}
\label{models}
\end{figure*}

\begin{figure*}
\begin{center}
\begin{tabular}{cc}
\includegraphics[width=7cm]{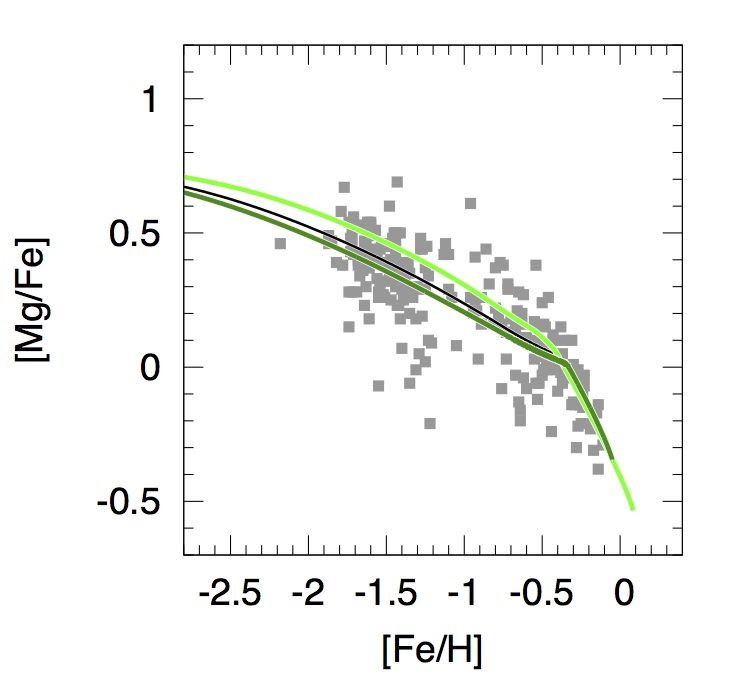} &
\includegraphics[width=7cm]{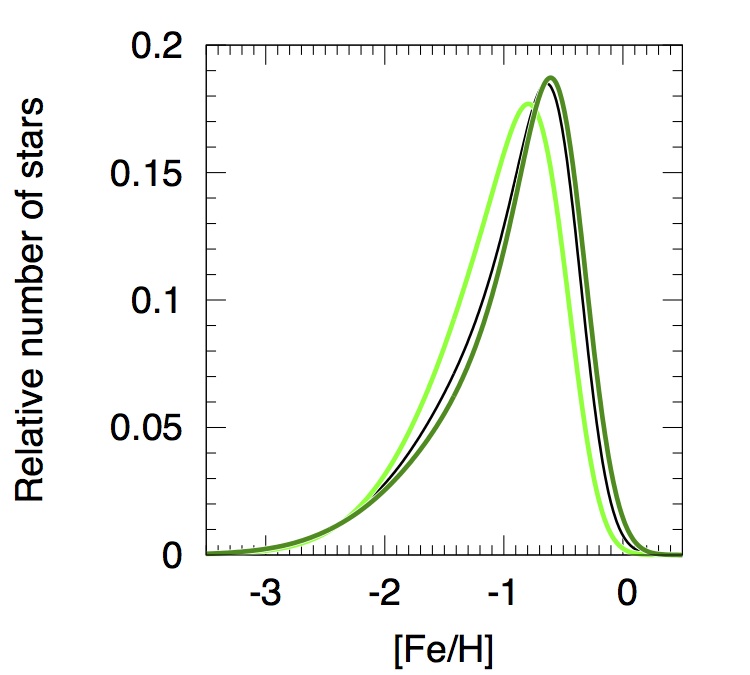} \\
\end{tabular}
\caption{The effect of a prolonged (dark green curves, gas stipping starting 3.5~Gyr ago) or shortened (light 
green curves, gas stripping starting 11.5~Gyr ago) star formation activity on the predicted [Mg/Fe] versus [Fe/H] 
behaviour (left-hand panel) and stellar metallicity distribution (right-hand 
panel) of Sagittarius stars.}
\label{mdf}
\end{center}
\end{figure*}

The stellar yields are from \cite{kar10} for low- and intermediate-mass stars, 
\cite{nom13} for massive stars and \cite{iwa99} for type Ia supernovae. The 
adopted stellar nucleosynthesis from massive stars is slightly updated with 
respect to our previous work, so, we checked that with our choice of the 
stellar yields the [X/Fe] versus [Fe/H] trends of elements from C to Zn in the 
Milky Way are reproduced. We run a model in which all massive stars explode as 
normal type II supernovae and a model in which all stars above 20~M$_\odot$ 
explode as hypernovae (solid and dotted lines, respectively, in 
Fig.~\ref{models}).

The values of the free parameters $M_{\rm{inf}}$, $\tau$, and $\nu$ are fixed in 
such a way that the chemical properties of Sgr stars are reproduced, while a 
stellar mass of $M_\star \simeq$~7\,$\times$10$^8$~M$_\odot$ at infall is 
obtained. 
For the comparison with the model predictions, we use data from this 
work for [Mg/Fe] and [Ca/Fe]. 
We define theoretical 
[$\alpha$/Fe] ratios as 0.5\,$\times$\,[Mg/Fe]~+ 0.5\,$\times$\,[Ca/Fe] for a 
twofold reason: first, we want to compare with the data presented in this work 
and, second, current yields of Ti still lead to theoretical [Ti/Fe] versus 
[Fe/H] trends that do not reproduce the one observed in the Milky Way 
\citep{rom10,kirby11,nom13}, thus making any comparison involving Ti not meaningful. 
For the iron-peak elements Mn, Cu, and Zn, we use data from \citet{c10b,sbo07}
and \citet{mcw13}. Measurements of Na and Al are also available for Sgr main 
body stars. However, for stars belonging to M54 these may reflect the peculiar 
enrichment processes occurring in a globular cluster, rather than the global 
galactic enrichment process. Therefore, we do not consider these elements in 
our study. On the other hand the [Mg/Fe] spread in M54 is moderate (C10b), thus 
it is not expected to seriously affect the general trend with [Fe/H]. 
The analysis of Eu data, instead, is complicated by the fact that this element, 
like other r-process elements, is likely to originate, at least partly, from 
compact binary mergers \citep{mat14}. Since we do not have an estimate of the 
rate of compact binary mergers in the progenitor of Sgr, we prefer not to deal 
with Eu in this paper.

From Fig.~\ref{models} it can be seen that our pure chemical evolution model, 
notwithstanding the rough assumptions it makes about the effects of the 
interaction with the Galaxy, is quite successful in reproducing the average 
trends of the analysed elements in Sgr. We stress that the abruptly decreasing trends of [Mg/Fe] and [Ca/Fe] for [Fe/H]~$> -$0.4 (red portions of the curves in 
Fig.~\ref{models}) can be obtained only by assuming that an efficient stripping is removing all the gas starting from 
7.5~Gyr ago; without it, a much milder decrease would be found, at variance with the observations. 
Comparing the predictions of a model for the whole Sgr system with data from the central nucleus alone may
not appear fully appropriate, but we verified that our model reproduces satisfactorily also the
[$\alpha$/Fe] versus [Fe/H] pattern found by \citet{deboer} in the Sgr Stream\footnote{Note that \citet{deboer}
adopt another definition for [$\alpha$/Fe], [$\alpha$/Fe]~= 0.5\,$\times$\,[Mg/Fe]~+ 0.3\,$\times$\,[Ti/Fe]~+ 
0.1\,$\times$\,[Ca/Fe]~+ 0.1\,$\times$\,[Si/Fe]. This is the reason why we do not include their data in Fig.~\ref{models}.}, in the metallicity range spanned by their data.

The model with normal supernovae is 
favoured, at least  for [Fe/H]~$> -$2.0, while the only very-metal poor stars seems to suggest that
hypernovae may have played a role at early epochs.
It would be interesting to test this hypothesis by measuring the abundance ratios in metal-poor stars for other 
elements, such as Zn or Mn, whose nucleosynthesis is strongly affected by the explosion energy. 
Furthermore, it is reasonable to assume that a multi-zone model 
would be better suited to describe the evolution of Sgr precursors, since the 
outer parts of the galaxy are expected to evolve more slowly than the inner parts. 
Indeed, \cite{deboer} find that the knee in the [$\alpha$/Fe] versus [Fe/H] 
relation for stream stars occurs at [Fe/H]~$\sim -$1.2 dex, while our main body 
sample points to a higher metallicity value of [Fe/H]~$\sim -$0.4 dex. Much 
more detailed (and homogeneous) chemical abundance data for a stream sample as 
large as that presented in this work are required before we can attempt this 
kind of modelling. Also, we remark that the constraints on the timescales of 
chemical enrichment and stripping in Sgr that can be set with arguments based 
on the trends of abundance ratios with metallicity alone are quite loose: by 
assuming a short-lasting, more efficient main episode of star formation from 14 
to 10.5~Gyr ago, with stripping starting 11~Gyr ago, or a long-lasting (star formation  
from 14 to 1 Gyr ago, with a drop in efficiency 2 Gyr ago\footnote{Note that this model is 
broadly consistent with the star formation scenario emerging from the analysis by \citet{siegel}.}), 
less efficient main episode of star formation and stripping starting 3.5~Gyr ago, we obtain almost 
the same [Mg/Fe] versus [Fe/H] theoretical trend (Fig.~\ref{mdf}, left panel). The stellar metallicity distribution function, 
instead, could be a better diagnostic for the epoch of gas-stripping, since its peak moves from 
[Fe/H]~=$-$0.85 to [Fe/H]~=$-$0.6 in the two extreme cases considered here. One would need accurate [Fe/H] 
measurements for a complete sample of Sgr stars to be able to discriminate among different scenarios from 
the metallicity distribution; however models with later gas-stripping produce distributions peaking 
at higher metallicity, in better agreement with the observed peak of the metal-rich component of Sgr. 
Finally, we note that our model assumes a \cite{krou02} IMF 
with a slope $x$~= 1.7, i.e. slightly steeper than Salpeter, as usually adopted in 
models of the solar neighbourhood region. All in all, it seems that a more 
exotic IMF strongly suppressing the high-mass stars, as proposed by MWM13, is not required in order 
to explain the available data.

\section{Summary and conclusions}
\label{conc}

The main results obtained from the analysis of 235 giant stars in the globular cluster 
M54 and the nuclear region of the Sagittarius dwarf galaxy, Sgr,N, are 
summarised as follows:
\begin{itemize}

\item{} We detect for the first time a metallicity gradient within the nucleus Sgr,N.
The peak of the metal-rich component shifts from [Fe/H]=$-0.38$ 
for $R\le 19.0$~pc to  [Fe/H]=$-0.57$ for 38.0~pc$<R\le 70.0$~pc (projected distance).
This evidence, together with the radial distribution of stars of different ages and metallicities in the
innermost $\simeq 20$~pc, indicates that the stars in the metal-rich component of Sgr,N  formed
outside-in over a few Gyr, with each subsequent generation of stars more centrally concentrated than 
the previous one.

\item{} Hence, the Sgr,N stellar population is not dynamically mixed and retains, in this very confined 
spatial region, some memory of its formation. We showed previously in B08 that the timescale 
for M54 to decay into the center of Sgr by dynamical friction is very short, on the order of 
approximately a Gyr. During this process, the globular cluster will have caused a massive wake 
in the ambient medium, which will have stirred the inner regions very effectively. 
It is unlikely that any pre-existing structure such as Sgr,N could have survived as a chemically 
inhomogeneous structure after this mixing. This opens up the fascinating possibility of being able 
to place an upper limit to the time since the arrival of M54 into the Sagittarius core by measuring 
the age of the Sgr,N stellar population. 
In the framework outlined by \citet{siegel}, this would suggest that the infall of M54 occurred more than $\sim$6 Gyr ago. 
Indeed, it is possible that this massive and compact system may have had a role in collecting the enriched 
gas that went on to form Sgr,N at the very center of the Sagittarius galaxy.

\item{} Assuming that M54 shares the same chemical pattern as Sgr~dSph stars in the metallicity
range spanned by its stars, we recover the [Mg/Fe], [Ca/Fe] and [Ti/Fe] vs. [Fe/H] trend
at the center of Sgr from [Fe/H]$\simeq -2.2$ to [Fe/H]$\simeq -0.2$. 

\item{} For [Fe/H]$< -1.0$
the $\alpha$-element abundance ratio in Sgr is the same as in the halo of the Milky Way and
in classical dSphs. On the other hand, at higher metallicity it is different from galaxy to galaxy:
Sgr being intermediate between the MW and Fornax dSph. 

\item{} In the $0.5\times$([Mg/Fe]+[Ca/Fe]) vs. [Fe/H] diagram Sgr stars follow a trend very similar to LMC stars. This
may suggest a similarity (e.g., in mass) between the progenitor of Sgr and the LMC, in good agreement with recent estimates of the Sgr mass at infall \citep{nied12,deboer,gibbons}. The comparison in the same plane with other dSphs corroborates this conclusion. 

\item{} We reproduce the main observed abundance trends of Sgr~dSph with a chemical evolution model implying 
an episode of strong gas loss occurring from 7.5~Gyr to 2.5 Gyr ago, presumably starting at the first peri-Galactic passage of the dwarf after its infall into the Milky Way. In our model the gas stripping event is required to reproduce the observed abundance trends in the most metal-rich regime ([Fe/H]$\ga -0.5$). However, because of the lack of an observed stellar metallicity distribution function for a complete sample of Sgr stars, the timescales for the occurrence of the stripping and chemical enrichment can be only poorly-constrained and the one presented here is only one among different possible evolutive scenarios.

 \end{itemize}
    
\begin{acknowledgements}

We are grateful to the anonymous referee for the useful comments and suggestions.
DR benefited from the International Space Science Institute (ISSI, Bern, CH), 
thanks to the funding of the team ``The Formation and Evolution of the Galactic Halo''.
Most of the data presented herein were obtained at the W.M. Keck Observatory, which is operated as a scientific partnership 
among the California Institute of Technology, the University of California and the National Aeronautics and Space Administration. 
The Observatory was made possible by the generous financial support of the W.M. Keck Foundation.
Partially based on Advanced Camera for Surveys (ACS) observations collected with the HST within the programme GO-10775.
This research has made use of the SIMBAD database, operated at CDS, Strasbourg, France.
This research has made use of the NASA/IPAC Extragalactic Database (NED) which is operated by the Jet Propulsion Laboratory, 
California Institute of Technology, under contract with the National Aeronautics and Space Administration. 
This research has made use of NASA's Astrophysics Data System.
\end{acknowledgements}

\bibliographystyle{apj}


\end{document}